\documentclass[referee]{raa}
\usepackage{graphicx,times}
\usepackage{natbib}
\usepackage{amssymb,amsmath}
\usepackage{rotating}
\usepackage{graphicx}
\usepackage{amsmath,amssymb}
\usepackage{longtable,lscape}
\usepackage{morefloats}

\begin{document}

   \title{A statistical study towards the high-mass BGPS clumps with the MALT90 survey
}

 \volnopage{ {\bf 2012} Vol.\ {\bf X} No. {\bf XX}, 000--000}
   \setcounter{page}{1}

\author{Xiao-Lan Liu\inst{1,2}, Jin-Long Xu\inst{1,2}, Chang-Chun Ning\inst{3},Chuan-Peng Zhang\inst{1,2}, Xiao-Tao Liu\inst{4}}

   \institute{National Astronomical Observatories, Chinese Academy of Sciences,
             Beijing 100012, China; {\it liuxiaolan@bao.ac.cn}\\
       \and
             NAOC-TU Joint Center for Astrophysics, Lhasa 850000, China \\
       \and  Tibet University, Lhasa, Tibet, 850000, China \\
       \and  Department of Finance, Central China Normal University, Wuhan 430079, China \\
 \vs \no
   {\small Received 2012 June 12; accepted 2012 July 27}
}

\abstract{ In this work, we perform a statistical investigation towards 50 high-mass clumps using the data from the Bolocam Galactic Plane Survey (BGPS) and the Millimetre Astronomy Legacy Team 90-GHz survey (MALT90). Eleven dense molecular lines (N$_2$H$^+$(1-0), HNC(1-0), HCO$^+$(1-0), HCN(1-0), HN$^{13}$C(1-0), H$^{13}$CO$^+$(1-0), C$_2$H(1-0), HC$_3$N(10-9), SiO(2-1), $^{13}$CS(2-1) and HNCO$(4_{4,0}-3_{0,3}))$ are detected. N$_2$H$^+$ and HNC are shown to be good tracers for clumps in virous evolutionary stages since they are detected in all the fields. And the detection rates of N-bearing molecules decrease as the clumps evolve, but those of O-bearing species increase with evolution. Furthermore, the abundance ratios [N$_2$H$^+$]/[HCO$^+$] and Log([HC$_3$N]/[HCO$^+$]) decline with Log([HCO$^+$]) as two linear functions, respectively. This suggests the transformation of N$_2$H$^+$ and HC$_3$N to HCO$^+$ as the clumps evolve. We also find that C$_2$H is the most abundant molecule with an order of $10^{-8}$. Besides, three new infall candidates G010.214-00.324, G011.121-00.128, and G012.215-00.118(a) are discovered to have large-scaled infall motions and infall rates in the magnitude of $10^{-3}$ M$_\odot$ yr$^{-1}$.
}

   \authorrunning{X.-L. Liu et al. }            
   \titlerunning{A statistical study towards the high-mass BGPS clumps with the MALT90 survey}  
   \maketitle

\section{introduction}
The fate of an individual star is determined by its mass and chemical composition at birth. A massive star has several ways to march towards death, based on its initial mass. Therefore, understanding the initial physical and chemical characteristics of the massive stars can promote the investigations on stellar structure and evolution. However, it is difficult to assure the exact initial properties of the stars just through the observations and the numerical simulations. We had better establish a right theory on star formation to infer the mass and chemistry when the star was born. For decades of research, the theory regarding the low-mass star formation is relatively mature: disc accretion and driven molecular outflows \citep{Shu87}. However, various formation scenarios for high-mass stars are widely debated in the literature \citep[e.g.][]{Zinnecker07,Beuther07,Tan14}. Core accretion \citep{McKee02,McKee03} and competitive accretion \citep{Bonnell01,Bonnell04} are two controversial theories underlying current studies \citep{He16}. The core accretion model, similar to the low-mass star formation, essentially presumes that the star's environment has no great impact on the evolution of the isolated core. The final mass of a massive star is dominated by the self-gravitation, which is assumed to be supported either by thermal pressure, turbulence or magnetic fields \citep{McKee02,McKee03}. This model emphasizes a direct link between core and star, and therefore the core mass function (CMF) will have a similar shape to the stellar initial mass function (IMF) \citep[e.g.][]{Motte98}. While for the competitive accretion model, the circumstellar environment plays a vital role in the mass of a star, since a set of cores in different masses competitively accrete materials from their surrounding envelopes. Most of the higher-mass cores fragment into sub-cores and majority of the sub-cores do not continue to accrete significantly such that their masses are set from the fragmentation process. Meanwhile, a few higher-mass cores continue to accrete and become the higher-mass stars \citep{Bonnell06}. Hence, the competitive accretion model can give an explanation for the full range and distribution of stellar masses \citep{Bonnell01}.

From the perspective of observations, the later competitive accretion scenario seems more suitable for the massive star formation since the cases of an individual massive star accompanying numerous low-mass stars around in a cluster were frequently detected \citep{Bressert12}. Each of these observations just represents a kind of evolutionary state at a moment, not a continuous forming process. Instead, a statistic study towards a sample consist of high-mass cores in different evolutionary stages can help us to deduce the various law of the physical and chemical features during the evolution of high-mass star formation. Adopting the Spitzer 3.6, 4.5, 8.0 and 24 $\mu$m images, \citet{Guzman15} classified the whole MALT90 survey clumps by visual inspection into four consecutive evolutionary stages: quiescent (pre-stellar), proto-stellar, HII region, and photo-dissociation region (PDR). In terms of the mid-IR characteristics, quiescent clumps seem dark at 3.6-24 $\mu$m. Proto-stellar clumps either contain a 24 $\mu$m point source or are associated with an ``extended green object" \citep[EGO;][]{Cyganowski08}. The unresolved 24 $\mu$m emission, mostly produced by relatively hot dust ($\geq40$ K) \citep{Faimali12,Zhang16}, indicates the existence of embedded stars or protostars, whereas the EGO results from the shocks generated by the molecular outflows. Both HII regions and photodissociation regions (PDRs) show extended 8 $\mu$m emission, which originates mainly from polycyclic aromatic hydrocarbons \citep[PAHs;][]{Watson08} at 8 $\mu$m bandpass.

In this paper, we carry out a statistical study towards 50 high-mass clumps using the continuum 1.1 mm data from BGPS, together with the molecular lines from MALT90, aiming at exploring the environment where the massive stars form. Given that the dust emission is optically thin in molecular clouds, the 1.1 mm data can be utilized to display the internal structures and derive the physical information within clumps such as the H$_2$ column densities, the dust temperatures, masses and volume densities. Alternatively, the optically thin spectra N$_2$H$^+$(1-0), HN$^{13}$C(1-0) and H$^{13}$CO$^+$(1-0) yet can trace the internal structures of clumps, the fragmentation of clumps and provide the velocity information of the observed sources, an important factor to derive the distances. While the optically thick lines HNC(1-0) and HCO$^+$(1-0) are good indicators of infall motions, implying the existence of star formation. Furthermore, we also research the chemical compositions, properties and evolutions by analyzing the molecular lines. With regard to the remaining parts of this paper, section 2 gives a description about the data archive and section 3 is about the results. The discussion is in section 4. We summarize our conclusions in section 5.

\section{Data and sample}
\subsection{Archival data}
The Bolocam Galactic Plane Survey \footnote{http://irsa.ipac.caltech.edu/data/$\rm BOLOCAM\_GPS$/} \citep[BGPS;][]{glenn09} is a 1.1 mm continuum survey of the Galactic Plane made using Bolocam \citep{Glenn03,Haig04} on the Caltech Submillimeter Observatory (CSO). This mm band traces the thermal dust emission, which can reveal the repositories of the dense molecular gases, ranging in scale from cores to entire clouds. Hence, the connection between these gases and the star formation regions may be explored. The BGPS has a full width at half-maximum (FWHM) effective beam size of $33''$ and totally covers 192 square degrees, including a blind survey of the inner Galaxy spanning from $\rm -10^\circ<l<90^\circ.5$ where $\rm |b|<0.5^\circ$, the Cygnus X spiral arm ($\rm 75.5^\circ \leq l \leq 87.5^\circ, \, |b|\leq 1.5^\circ$), cross-cuts ($\rm l= 3^\circ,\,15^\circ,\,30^\circ,\, 31^\circ, \, |b|\leq 1.5^\circ$), and four targeted regions in outer Galaxy which are IC1396 (9 square degrees, $\rm 97.5^\circ \leq l \leq 100.5^\circ, \, 2.25^\circ \leq l \leq 5.25^\circ$), a region towards the Perseus Arm (4 square degrees centered on $\rm l = 111$, $\rm b=0$ near NGC7538), W3/4/5 (18 square degrees, $132.5^\circ \leq l \leq 138.5^\circ$), along with Gem OB1 (6 square degrees, $\rm 187.5^\circ \leq l \leq 193.5^\circ$). \citet{Rosolowsky10} performed a source catalogue of this survey, which consists of 8358 sources and is 98$\%$ complete from 0.4 Jy to 60 Jy over all object sizes for which the survey is sensitive ($<3'.5$).

The Millimetre Astronomy Legacy Team 90-GHz (MALT90) Survey \citep{Jackson13} is a large international project conducted with the ATNF Mopra 22-m telescope, which simultaneously images 16 molecular lines near 90 GHz with the On-the-Fly (OTF) mapping mode. These dense molecular lines characterize the physical and chemical conditions of the high-mass star formation clumps in different evolutionary stages (pre-stellar, proto-stellar, HII, and PDR). This survey covers the Galactic longitude ranges $\rm 300^\circ <l< 357^\circ$ (4th quadrant) and $\rm 3^\circ <l< 20^\circ$ (1th quadrant), with the effective angular and spectral resolution as $\sim 36''$ and $\sim$0.11 km s$^{-1}$, respectively. The final data were recorded in the antenna temperature scale of $\rm T_A^\ast$ (K) and the sensitivity ($\rm T_A^\ast \times {rms}$ at 0.11 km s$^{-1}$) is about 0.2 K. The conversion for the line intensities to the main beam brightness temperature scale is made using the formula $\rm T_{MB}=T_A^\star /\eta_{MB}$, where $\rm\eta_{MB}$ is the main beam efficiency about 0.49 at 86 GHz and 0.44 at 110 GHz \citep{Ladd05}. Extrapolation using the Ruze formula gives the $\rm\eta_{MB}$ values in the range of 0.49-0.46 for the 86.75-93.17 GHz frequency range of MALT90. The MALT90 data were obtained from the online archive \footnote{http://atoa.atnf.csiro.au/MALT90} and reduced by the software CLASS (Continuum and Line Analysis Single-Dish Software) and GREG (Grenoble Graphic) \footnote{http://www.iram.fr/IRAMFR/GILDAS}.

The Spitzer IRAC 8 $\mu m$, MIPS 24 $\mu m$, Herschel PACS 70, 160 $\mu m$ together with SPIRE 250, 350, 500 $\mu m$ images were downloaded from the Galactic Legacy Infrared Mid-Plane Survey Extraordinaire (GLIMPSE), a 24 and 70 Micron Survey of the Inner Galactic Disk with MIPS (MIPSGAL) \footnote{http://irsa.ipac.caltech.edu/data/SPITZER/GLIMPSE} and the Herschel Science Archive (HSA)\footnote{http://irsa.ipac.caltech.edu/applications/Herschel} \citep{Poglitsch10,Griffin10}.

\subsection{Source selection and classification}
In order to characterize the high-mass clumps, the sample used in this study is a sub-sample from the BGPS catalogue, satisfying following points: (a) all the sources in the sample are covered in the survey region of MALT90; (b) the equivalent radius of each BGPS clump is at least twice beam size of MALT90 (about 36''), for the sake of reducing the beam dilution and distinguishing two cores within the same clump; (c) each BGPS clump in the sample has at least one molecular line detected whose signal-to-noise (S/N) is at or above the $3\sigma$ limit. As a result, 48 BGPS clumps are chosen first. Then seeing the N$_2$H$^+$(1-0) emission in Fig. ~\ref{fig7}, we find that clumps G12.215-00.118 and G354.00+00.474 both contain two distinguished sub-structures ( G12.215-00.118(a) and G12.215-00.118(b), G354.00+00.474(a) and G354.00+00.474(b)). It means 50 clumps finally identified. Table~\ref{tab1} presents some basic information for the 50 clumps from the BGPS catalog (ID, clump name, longitude and latitude of the peak, deconvolved angular radius, $40''$ aperture flux density and integrated flux density). However, the peaks of G12.215-00.118(a), G12.215-00.118(b), G354.00+00.474(a) and G354.00+00.474(b) in Table~\ref{tab1} correspond to the peaks of the N$_2$H$^+$ emission within them (the blue ``+" marked in Fig. ~\ref{fig7}) and offset from the previous ones measured by \citet{Rosolowsky10}. Meanwhile, referring to the classification towards the whole MALT90 survey clumps of \citet{Guzman15}, we find that our sample comprises 2 PDR, 25 HII regions, 19 proto-stellar, 3 per-stellar and 1 uncertain (G12.215-00.118(a), G12.215-00.118(b), G354.00+00.474(a) and G354.00+00.474(b) are reclassified according to the method of \citet{Guzman15} with the Spizter 3.6, 4.5, 8.0 and 24 $\mu$m images. G12.215-00.118(a), G354.00+00.474(a) and G354.00+00.474(b) are HII regions, but G12.215-00.118(b) is in an earlier stage of proto-stellar). Column 16 of Table~\ref{tab1} lists the evolutionary stages of all clumps in our sample. It is valid that we integrate the PDR clumps into HII regions as \citet{hoq13} had done before and our PDR clumps are only two. In short, the selected sample encompasses three different evolutionary stages from pre-stellar(3), proto-stellar(19) to HII/PDR(27).

\section{Results}
\subsection{Detected species, distances and dust temperatures}
\begin{figure}
  \centering
\includegraphics[angle=0,scale=0.40]{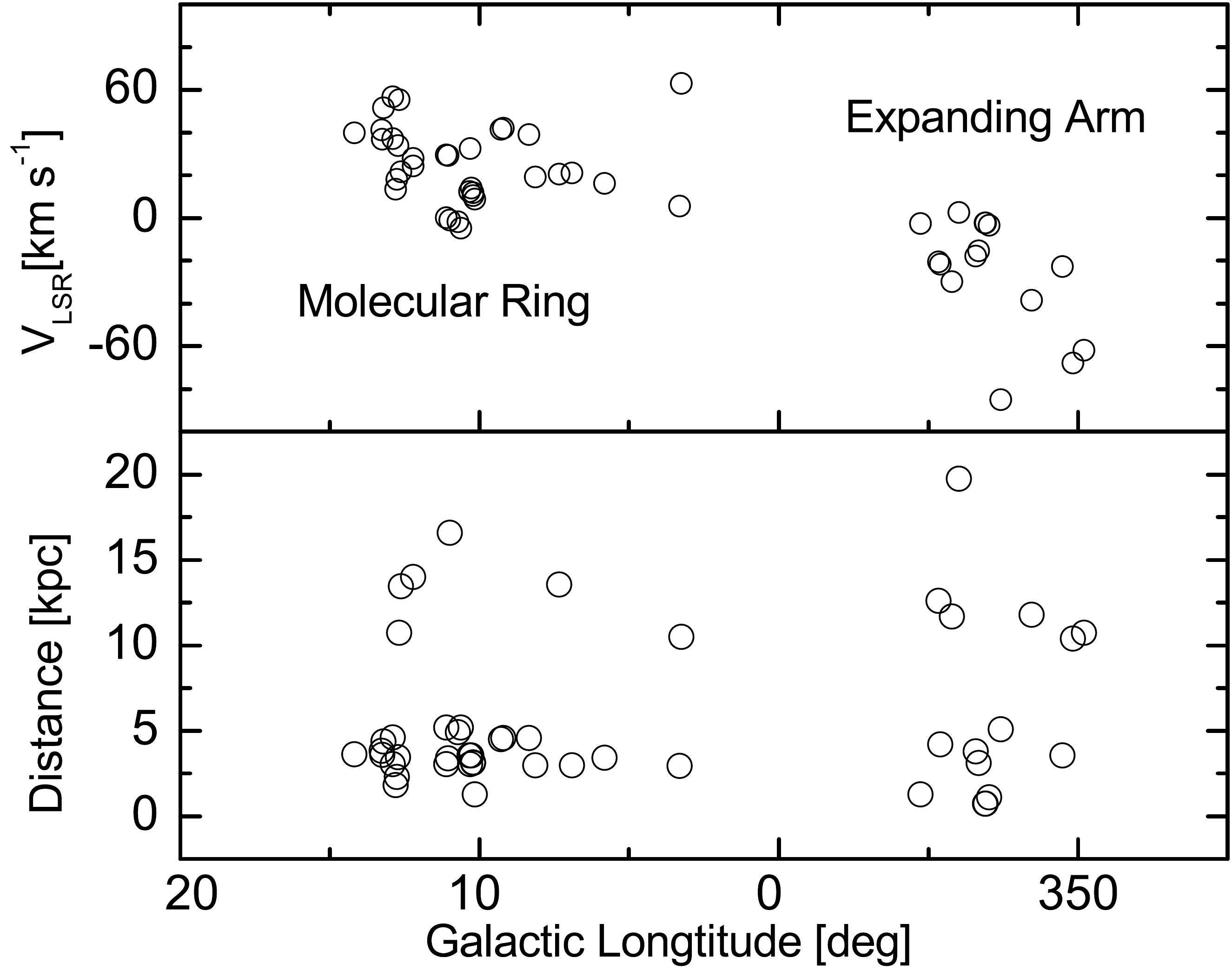}
\caption{Upper panel: the systemic velocities derived from the gaussian fittings towards N$_2$H$^+$(1-0) lines vs. Galactic longitude distribution towards 50 clumps. Lower panel: the distances vs. Galactic longitude distribution towards 50 clumps. }
\label{fig1}
\end{figure}

\begin{figure}
 \begin{minipage}[t]{0.5\textwidth}
  \centering
   \includegraphics[angle=0,scale=1.1]{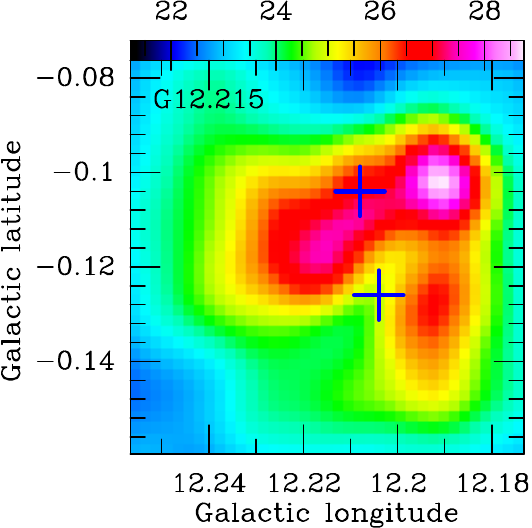}
  \end{minipage}%
  \begin{minipage}[t]{0.5\textwidth}
  \centering
   \includegraphics[angle=0,scale=1.1]{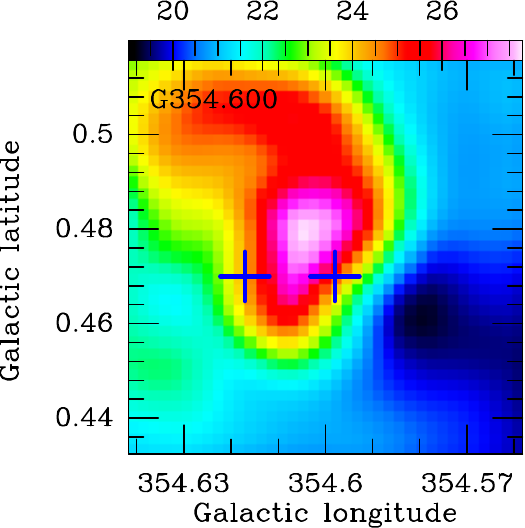}
  \end{minipage}%

\caption{Maps of the dust temperature $\rm T_{dust}$ towards G12.215-00.118 (Left) and G354.600+00.474 (Right) built on the SED fitting pixel by pixel. The ``$+$'' in each panel mark the peak positions of the sub-structures.}
\label{fig2}
\end{figure}

Table~\ref{tab2} shows the detected molecular lines of MALT90 and their detection rates. We find eleven spectra detected. N$_2$H$^+$(1-0) and HNC(1-0) are observed towards all the fields. This suggests that these two species are good indicators for tracing the whole procedure of the high-mass star formation. And the N$_2$H$^+$(1-0) lines can detect the velocity information and the internal structures since they are likely to be optically thin in the molecular clouds. Additionally, the detection rates of HCO$^+$(1-0), HCN(1-0) and HN$^{13}$C(1-0) are declining eventually with the evolutionary stages, but those of H$^{13}$CO$^+$(1-0), C$_2$H(1-0), HC$_3$N(10-9) present upward tendencies. This implies the chemical evolutions of Nitrogen-bearing, Oxygen-bearing and Carbon-bearing species. Furthermore, we find five clumps with SiO emission, excited by the shocks from the outflows. HNCO$(4_{4,0}-3_{0,3})$ is usually used to trace the hot core and shock-chemistry. In this work it is detected only in a pre-stellar clump G003.254+00.410, close to the Galactic center. And only clump G353.412-00.360 presents the $^{13}$CS(2-1) emission.

Distances of 30 clumps have been obtained from the literature. For the remaining 20 sources, combining the radial velocities of N$_2$H$^+$(1-0) (see Table~\ref{tab3} column 6), we can derive the distances of them using the Bayesian Distance Calculator \footnote{http://bessel.vlbi-astrometry.org/bayesian}\citep{Reid16}, which leverages the results to significantly improve the accuracy and reliability of distance estimates to other sources known to follow spiral structure. This calculator is proved to be more reliable by \citet{Paron13} through a comparison that the well-known distances of several HII regions and one SNR are close to the results calculated from the Bayesian Distance Calculator. Therefore, it is utilized in this work. The derived distances for the 50 clumps are listed in Table~\ref{tab1} column 5. Moreover, Fig.~\ref{fig1} shows the systemic velocity (V$\rm_{LSR}$ of N$_2$H$^+$(1-0) from section 3.4) and the distance distributions of 50 clumps along the Galactic longitude. Obviously, the sources in our sample concentrate on two active star formation regions-the Molecular Ring (1st quadrant ) and the Expanding Arm (4st quadrant).

On the other hand, the dust temperatures of the objects in our sample are taken from \citet{Guzman15}, except for G12.215-00.118(a), G12.215-00.118(b), G354.600+00.474(a) and G354.600+00.474(b), because they deviate from the original peaks of BGPS G12.215-00.118 and G354.600+00.474. We derive their temperatures by ourselves based on the pixel-by-pixel spectral energy distribution (SED) method provided by \citet{Wang15}. The high-quality Hi-GAL data covering a widespread wavelength (70-500 $\mu$m) can be utilized to investigate the dust properties of the entire clouds. Following the steps of \citet{Wang15}, firstly we should remove the background. Fourier transfer (FT) can transform the original images into Fourier domain and separate low and high spatial frequency components. The low-frequency component represents large-scale background and foreground emission, while the high-frequency maintains the emission of interest. Then the inverse FT backs them into image domain separately. Detailed illustrations of this FT method can be found in \citet{Wang15}. After removing the background and foreground emission, we convolve the images to the same resolution $45''$ and rebin them to the same pixel size $11.''5$, corresponding to the measured beam and pixel of Hi-GAL observations at 500 $\mu$m \citep{Traficante11}.

For each pixel, the intensity varies with wavelength as
 \begin{equation}
\rm  S_\nu = B_\nu (1-e^{-\tau_\nu}),
\label{eq1}
\end{equation}
in which Planck function B$_\nu$ is modified by optical depth
 \begin{equation}
\rm  \tau_\nu = \mu_{H_2}m{_H}\kappa_\nu N_{H_2}/ R_{gd},
\label{eq2}
\end{equation}
here $\rm \mu_{H_2} =2.8$ is the mean molecular weight considering the contributions of He and other heavy elements to the total mass \citep{Kauffmann08}. $\rm m_{H}$ is the mass of a hydrogen atom, $\rm N_{H_2}$ is the column density, and $\rm R_{gd}=100$ is the gas to dust ratio. $\kappa_\nu$ is the dust opacity per gram of gas, expressed as power law of frequency \citep{Ossenkopf94}
\begin{equation}
\rm  \kappa_\nu = 5.0 (\frac{\nu}{600\,GHz})^\beta cm^2 g^{-1},
\label{eq3}
\end{equation}
where the dust emissivity index $\beta$ has been fixed to be 1.75 \citep{Battersby11}. Therefore, the free parameters are the dust temperature T$\rm_{dust}$ and the H$_2$ column density $\rm N_{H_2}$.

\begin{figure}
  \centering
\includegraphics[angle=0,scale=0.40]{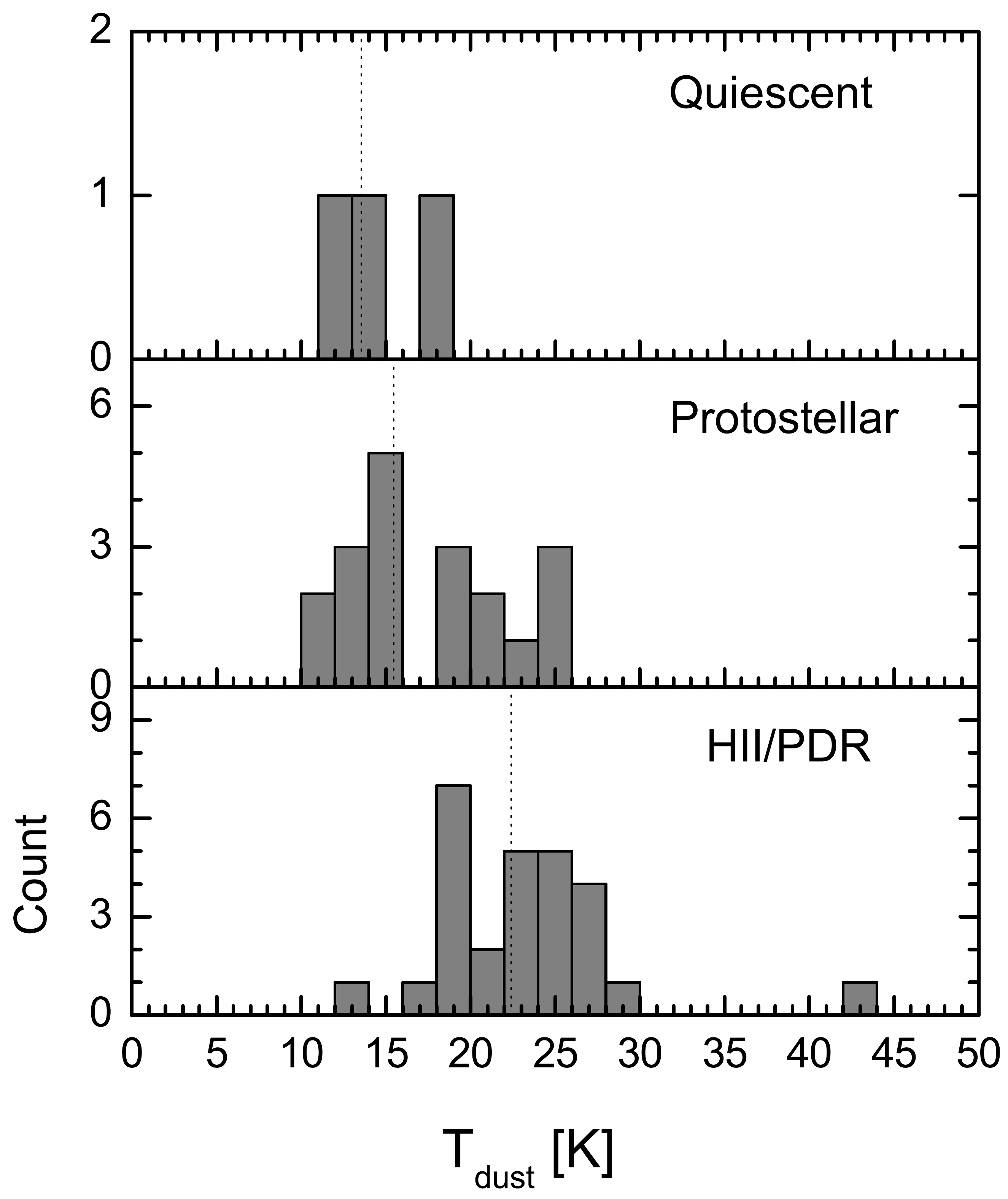}
\caption{The dust temperature distributions in each evolutionary stage. The dashed vertical lines indicate the median
values.}
\label{fig3}
\end{figure}

Figure~\ref{fig2} shows the temperature distributions towards G12.215-00.118 and G354.600+00.474. The averaged dust temperatures of G12.215-00.118(a), G12.215-00.118(b), G354.600+00.474(a) and G354.600+00.474(b) are 26.9, 25.1, 24.8 and 24.1 K, respectively. All the temperatures are summarized in column 7 of Table~\ref{tab1}. Fig.~\ref{fig3} presents the histogram distributions of T$\rm_{dust}$. We find the typical ranges of T$\rm_{dust}$ in pre-stellar, proto-stellar, and HII/PDR stages as $\sim$ 10-12, 14-20 and 18-24 K. The median values are 13.5, 15.5 and 22.5 K respectively. It suggests that the temperatures within the clumps increase as they evolve.

\subsection{Clump masses and densities}

\begin{figure}
  \centering
\includegraphics[angle=0,scale=0.40]{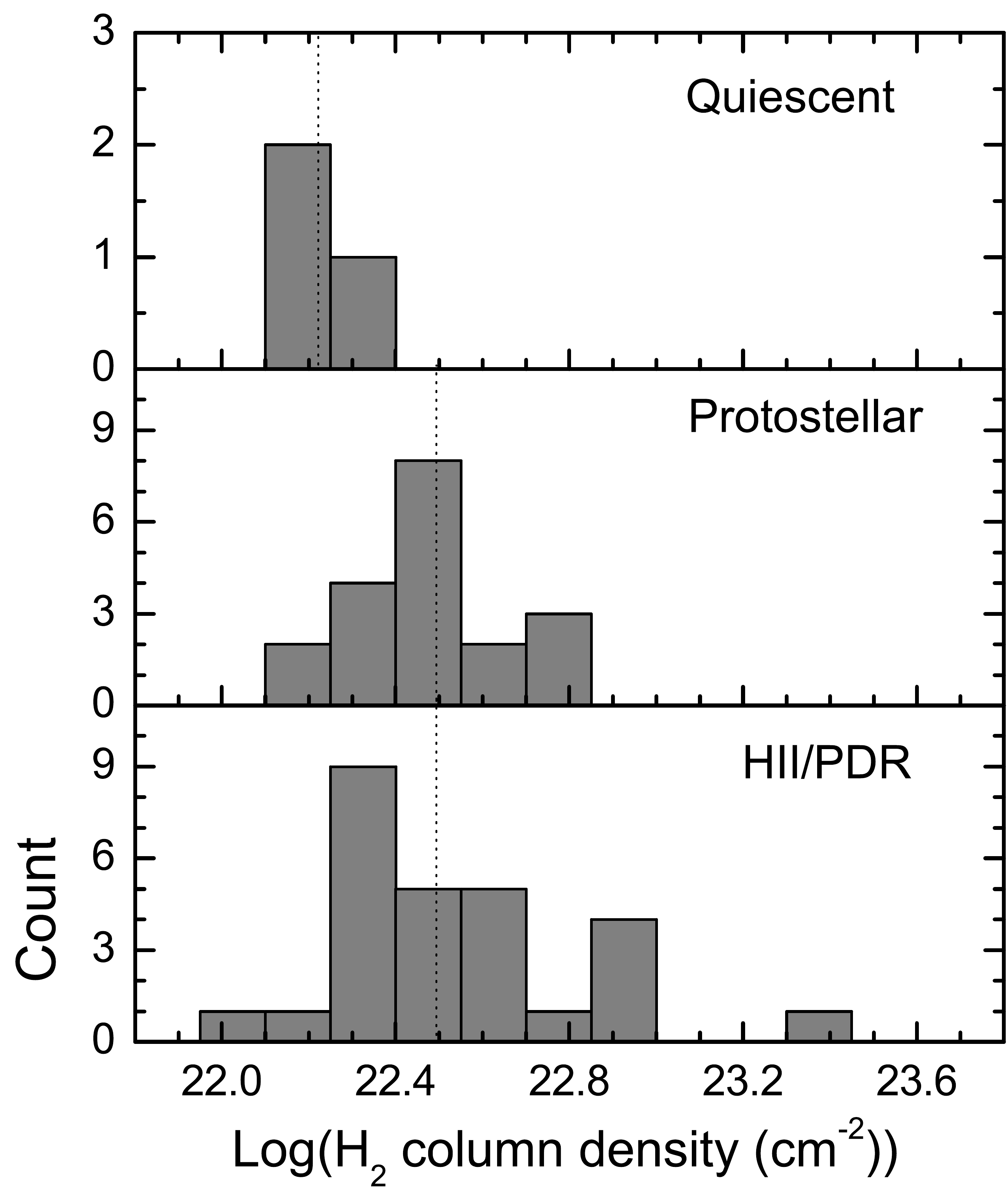}
\caption{H$_2$ column density histograms in each evolutionary stage. The dashed vertical lines indicate their median
values.}
\label{fig4}
\end{figure}

We calculate the beam-averaged H$_2$ column densities (N$\rm_{H_2}^{beam}$) by the following formula
\begin{equation}
\rm N_{H_2}^{beam}=\frac{S_\nu(40'')}{\Omega(40'') \mu_{H_2} m_H \kappa_\nu B_\nu(T_{dust}) R_{gd}} ,
\label{eq4}
\end{equation}
where S$_\nu$($40''$) is the flux density in an aperture of diameter $40''$, listed in column 10 of Table~\ref{tab1}. $\Omega(40'')= 2.95\times10^{-8}$ sr is the solid angle of the $40''$ aperture. $\rm\kappa_\nu =0.01114\, cm^2 \, g^{-1}$ at 271.1 GHz is the dust opacity per gram of gas and dust interpolated from equation~(\ref{eq3}). B$\rm_\nu(T_{dust})$ is the Planck function evaluated at $\rm T_{dust}$. Meanwhile, we apply an aperture correction of 1.46 to S$_\nu$($40''$) due to the sidelobes of CSO beam \citep{Aguirre11}. Here, we use S$_\nu$($40''$) as a measure of the flux within a beam since the solid angle of the $40''$ aperture is very close to the solid angle of the $33''$ effective beam ( $\Omega(33'')= 2.9\times10^{-8}$ sr ). And for G12.215-00.118(a), G12.215-00.118(b), G354.600+00.474(a) and G354.600+00.474(b), we also remeasure their radii and the integrated 1.1 mm fluxes. From Table~\ref{tab1}, we find the radii of G12.215-00.118(b), G354.600+00.474(a) and G354.600+00.474(b) are smaller than $40''$ aperture, while that of G12.215-00.118(a) is around $40''$. Therefore, we substitute their integrated fluxes for S$_\nu$($40''$). $\Omega(40'')$ is still used for them because of the beam dilution. The obtained $\rm N_{H_2}^{beam}$ are listed in column 12 of Table~\ref{tab1}. Fig.~\ref{fig4} presents the histograms of $\rm N_{H_2}^{beam}$ in pre-stellar, proto-stellar and HII/PDR stages, respectively. We find that the median values of H$_2$ column density are $(1.7\pm0.2)\times10^{22}$, $(3.1\pm0.2)\times10^{22}$ and $\rm (3.1\pm0.4)\times10^{22}\,cm^{-2}$ from pre-stellar clumps to HII/PDR. The above result indicates that the peak H$_2$ column density increases from pre-stellar to proto-stellar and HII/PDR. It suggests that clumps are eventually accumulating more materials as they evolve. However, the nearly equal median N$\rm_{H_2}$ from proto-stellar to HII/PDR might be attributed to the eventually strengthened feedbacks in the later evolutionary stage which are likely to disrupt the parent clumps.

\begin{figure}
  \centering
\includegraphics[angle=0,scale=0.40]{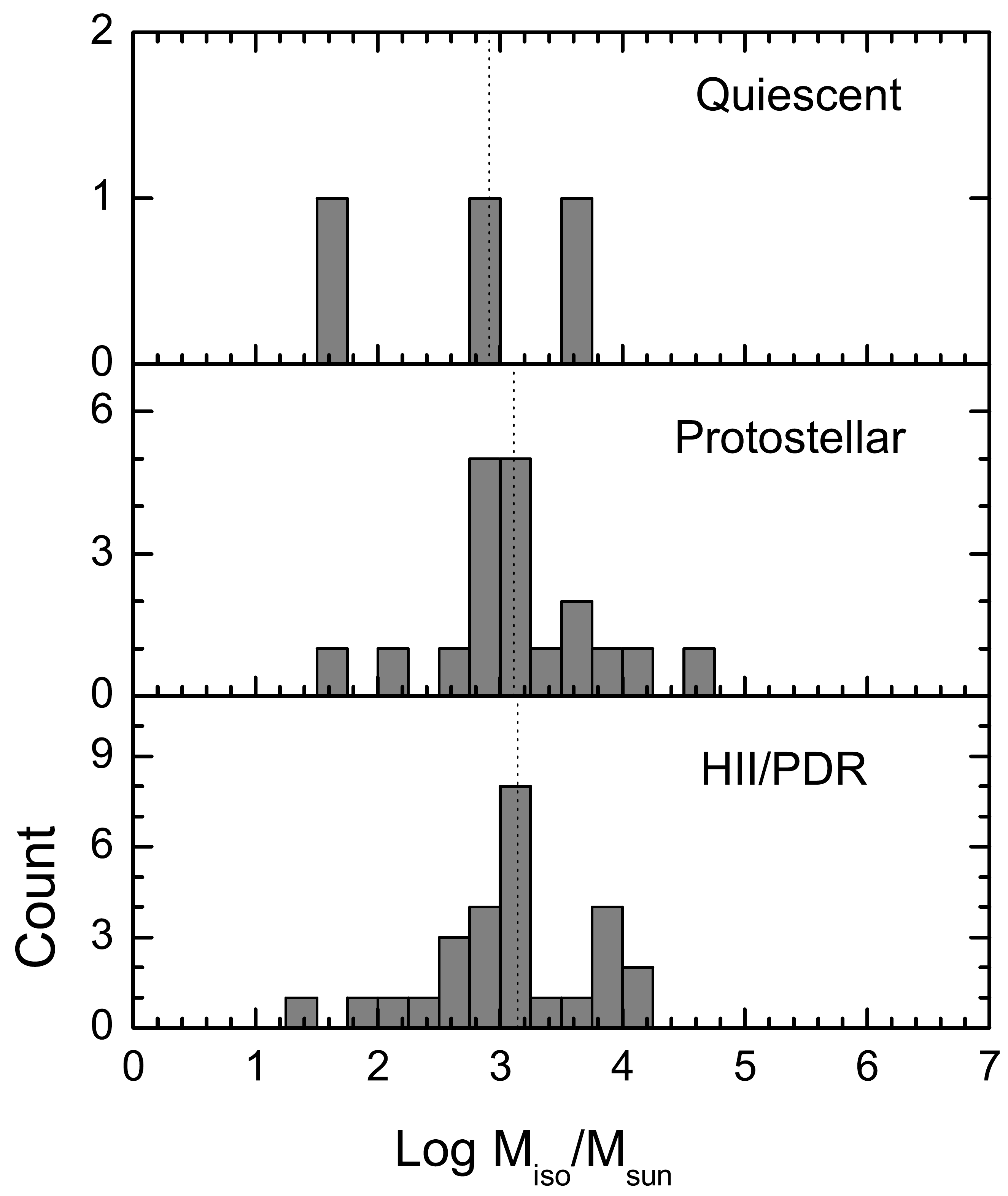}
\caption{The clump mass distributions in each evolutionary stage. The dashed vertical lines indicate their median values.}
\label{fig5}
\end{figure}

The isothermal dust masses, M$\rm_{iso}$, can be derived by the integrated 1.1 mm flux of the whole source via the expression
\begin{align}
\rm M_{iso} = \frac{S_{\nu}D^{2}}{R_{bg}\kappa_{\nu}B_{\nu}(T_{dust})}=13.1\ \mbox{M$_\odot$}\ (\frac{S_{\nu}}{1\ Jy})(\frac{D}{1\ kpc})^2(\frac{e^{13.0\ K / T_{dust}}-1}{e^{13.0\ K / 20.0\ K}-1}),
\label{eq5}
\end{align}
where S$_\nu$ is the integrated 1.1 mm flux. D is the distance. We plot the histograms of M$\rm_{iso}$ in various evolutionary stages in Fig.~\ref{fig5}. The median value per evolutionary stage (from pre-stellar, proto-stellar to HII/PDR) is $818\pm72$, $1155\pm83$ and $1220\pm88$ M$_\odot$, increasing slightly as a function of the evolutionary stage. This supports the possibility that clumps accumulate material continuously and efficiently with their evolution. Besides, the mass distributions of clumps in different stages are eventually concentrated on $\rm\sim 10^3\, M_\odot$ as they evolve.

\begin{figure}
  \centering
\includegraphics[angle=0,scale=0.50]{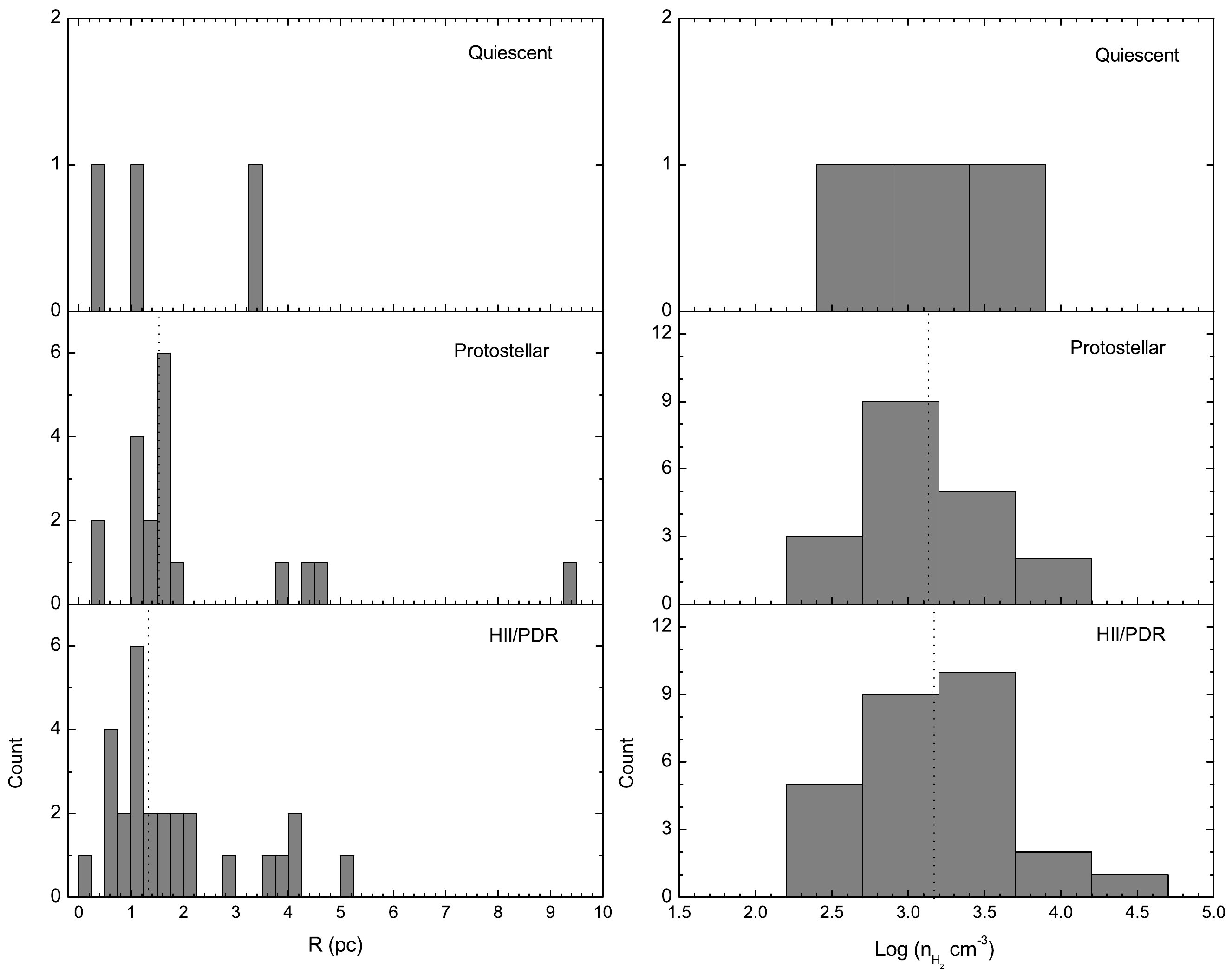}
\caption{Left panel: histograms of the effective radius R and the average H$_2$ $\bar{n}$ of clumps in stages from pre-stellar to HII/PDR. The dashed vertical lines indicate their median values.}
\label{fig6}
\end{figure}

Furthermore, we calculate the effective radius R and the average H$_2$ volume density $\rm \bar{n}$ according to the below relations
\begin{equation}
\rm R\simeq D\Theta,
\label{eq6}
\end{equation}

\begin{equation}
\rm  \bar{n} = 3M_{iso}/(4\pi R^3 \mu_{H_2} m_H),
\label{eq7}
\end{equation}
where $\Theta$ is the effective angle radius displayed in Table~\ref{tab1}, and  M$\rm_{iso}$ is the clump mass as calculated in equation~(\ref{eq5}). The final results of R and $\rm \bar{n}$ are summarized in columns 9 and 13 of Table~\ref{tab1}. Fig.~\ref{fig6} shows the effective radius and the average H$_2$ volume density distributions. From Fig.~\ref{fig6}, we find that the median radius from proto-stellar sources (1.5 pc) to HII/PDR clumps (1.3 pc) presents an inward contraction tendency, while the average H$_2$ volume density shows an increase with evolution. The median $\rm \bar{n}$ values from pre-stellar to proto-stellar to HII/PDR clumps are $(1.36\pm0.12)\times10^3$, $(1.41\pm0.11)\times10^3$, and $(1.45\pm0.11)\times10^3$ cm$^{-3}$, respectively. Both characteristics seem to cater for the evolutional tendency that the clumps will become denser and more compact.

\begin{figure}
 \begin{minipage}[t]{0.5\textwidth}
  \centering
   \includegraphics[angle=0,scale=0.8]{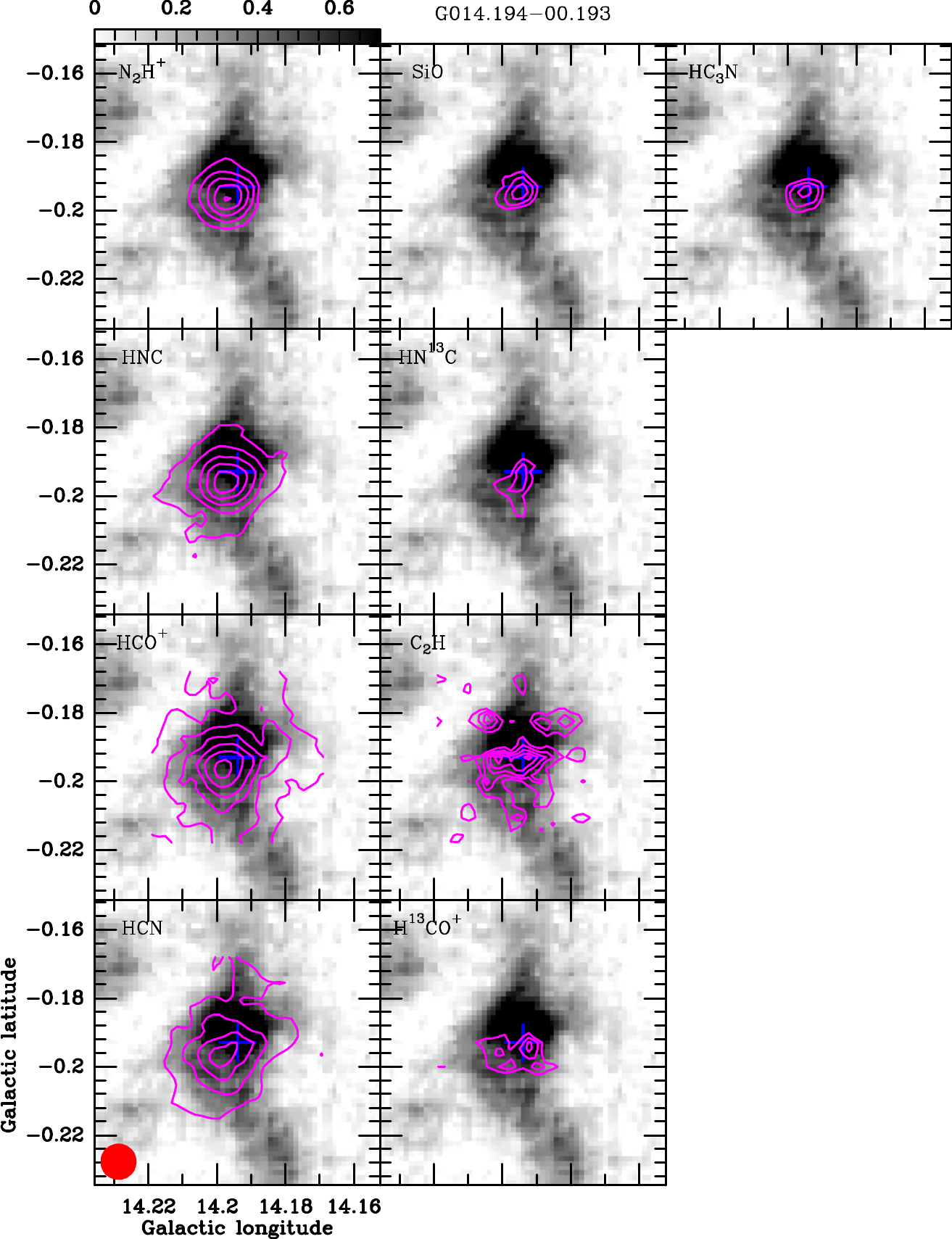}
  \end{minipage}%
  \begin{minipage}[t]{0.75\textwidth}
  \centering
   \includegraphics[angle=0,scale=0.9]{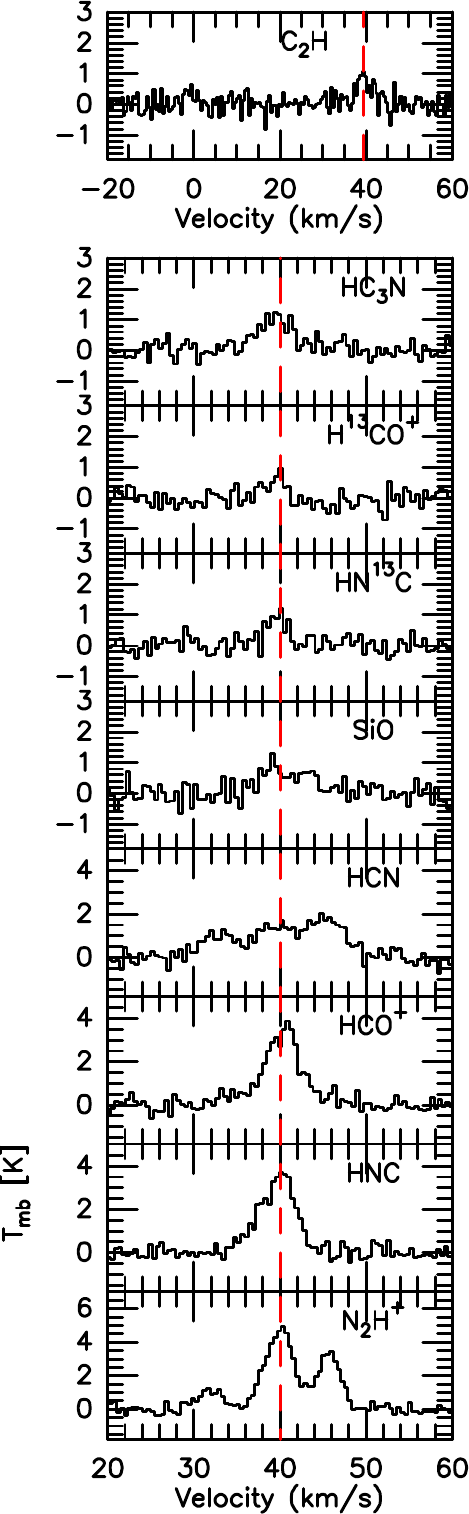}
  \end{minipage}%

\caption{Example of G014.194-00.193. Left-panel: The overlay maps of the BGPS 1.1 mm continuum emission (grey-scale) with the integrated intensity contours of the detected species which are denoted in the left-top corners. The colour bar indicates the flux with units Jy beam$^{-1}$ for 1.1 mm. The magenta contou levels start at 3$\sigma$ in steps of 0.8$\sigma$ ($\sigma$=1.29, 1.01, 0.94, 1.03, 0.55, 0.3, 0.33, 0.39, 0.56 K km s$^{-1}$) in the sequence of left-top to left-bottom then to right-top to right-bottom sub-figures. The symbols ``$+$'' marks the peak positions in Table~\ref{tab1}. Right-panel: The spectra are extracted from the peak position and the red dashed line marks the centroid velocity determined by the gaussian fitting to the N$_2$H$^+$(1-0) line. The other figures are in the supplementary.}
\label{fig7}
\end{figure}

\begin{figure}
  \centering
\includegraphics[angle=0,scale=0.40]{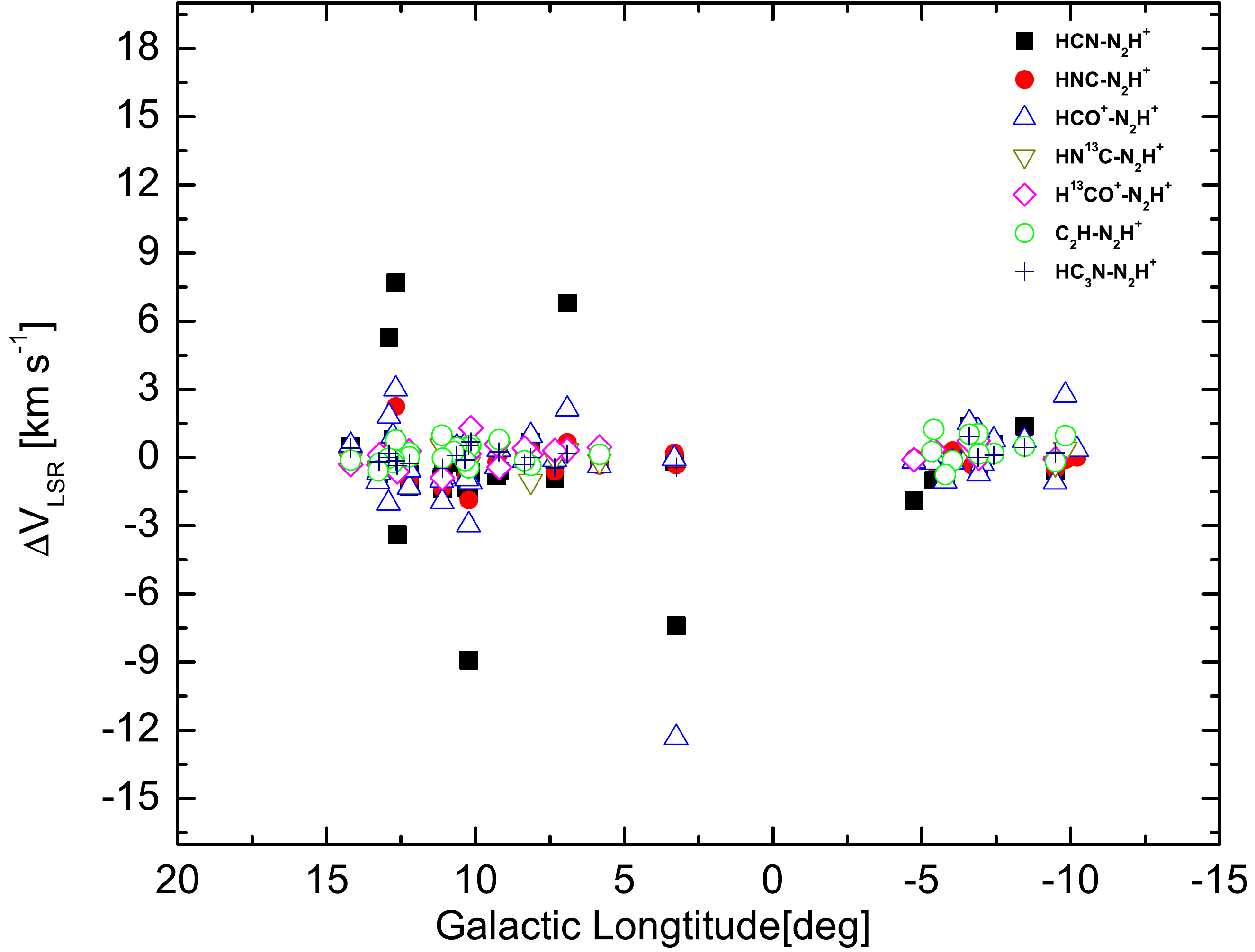}
\caption{The centroid velocity differences between N$_2$H$^+$(1-0) and other lines.}
\label{fig8}
\end{figure}

\subsection{Spatial distributions of the spectral-line emission}
In Table~\ref{tab2}, 11 out of 16 molecular lines of MALT90 and N$_2$H$^+$(1-0) are detected in all the 50 clumps. So firstly we employ this line to determine the velocity information (the centroid velocity and the integrated velocity interval) and the internal structure associated with each clump. The match between the N$_2$H$^+$(1-0) emission and the 1.1 mm continuum image for every clump provides us the integrated velocity range of each clump, shown in Table~\ref{tab3}. Based on these velocity ranges, we make the integrated intensity maps of the detected spectral lines towards the clumps in this subsection. We didn't use the 0th moment maps or the integrated intensity maps included in the MALT90 data archive, since the integrated velocity interval of each clump did not exactly match the clump. We remake the 0th moment maps and upgrade their corresponding uncertainties. The typical 1$\sigma$ error was found to be $\sim 0.5-2$ K km s$^{-1}$ for the detected lines.

The re-made 0th moment maps are presented in Fig.~\ref{fig7}, where the magenta contours indicating the molecular emission are overlaid on the BGPS 1.1 mm continuum data. The blue ``+" mark the peaks denoted in columns 2-3 of Table~\ref{tab1}. The contour levels begin at least 3$\sigma$, in order to get rid of the artificial signal and to illustrative purposes. Fig.~\ref{fig7} only shows the detected line emission, associated with the BGPS 1.1 mm continuum emission. These figures demonstrate that all the clumps are a single-core structure, except G12.215-00.118 and G354.600+00.474. They exhibit a two-core structure. It can be explained by the nearly equal effective beam sizes between BGPS 1.1 mm data ($33''$) and the MALT90 survey ($36''$) or the possibility that the spatial resolution of the telescope is not enough. So we need higher spatial resolution data to further investigate the fragmentation within the clumps. Additionally, the emission of N$_2$H$^+$(1-0), HNC(1-0), HCO$^+$(1-0) and HCN(1-0) are more extended than other species. And as the clumps evolve, the compositions of the dense gases will be richer and more complicated.

\subsection{Spectra lines and their physical parameters}
Figure~\ref{fig7} also shows the molecular lines extracted from the peak positions of the clumps, respectively. In terms of the spectral profiles, the hyperfine structures (hfs) of N$_2$H$^+$(1-0), HCN(1-0) and C$_2$H(1-0) are detected. However, near 90 GHz, N$_2$H$^+$(1-0) should be split into 15 hyperfine structures, out of which seven have a different frequency \citep{Pagani09,Keto10}. And C$_2$H(1-0) presents six hyperfine structures \citep{Reitblat80,Padovani09,Spielfiedel12}. Here, we find that seven components of N$_2$H$^+$(1-0) have blended into three components with a relative intensity of $1:5:3$. Meanwhile, two hyperfine structures ($\rm N_{J,F} = 1_{3/2,2}-0_{1/2,1}$ and $\rm N_{J,F} = 1_{3/2,1}-0_{1/2,0}$) of C$_2$H(1-0) are clearly observed in our MALT90 data. We make the gaussian fittings to all the detected spectra and the fitting parameters are listed in Table~\ref{tab3}. The hfs fittings are not done to N$_2$H$^+$(1-0), HCN(1-0) and C$_2$H(1-0), since more than 70$\%$ N$_2$H$^+$(1-0) exhibit an optical depth of 0.1 and over half of the later two lines present bad fittings. The physical parameters displayed in Table~\ref{tab3} are for the main components N$_2$H$^+$ ($\rm N_{J,F} = 1_{2,3}-0_{1,2}$), HCN ($\rm N_{J,F} = 1_{1,0}-0_{0,1}$) and C$_2$H ($\rm N_{J,F} = 1_{3/2,2}-0_{1/2,1}$), respectively. The systemic velocity of each clump is marked by the red dashed lines in Fig.~\ref{fig7}. In order to find out other indicators of systemic velocity, we plot the centroid velocity differences between other molecular lines and N$_2$H$^+$(1-0) in Fig.~\ref{fig8}. From Fig.~\ref{fig8}, we realize that the centroid velocities of all the molecular lines appear to coincide with those of N$_2$H$^+$(1-0), except HCO$^+$(1-0) and HCN(1-0). For HCO$^+$(1-0), the V$\rm_{LSR}$ of clump G003.254+00.410 shows an up to 12 km s$^{-1}$ red-shift and the remaining clumps present a difference within 3 km s$^{-1}$. For HCN(1-0), the V$\rm_{LSR}$ of seven clumps in the 1th quadrant of Galaxy emerge great gaps (3-12 km s$^{-1}$) compared with those of N$_2$H$^+$(1-0). The great red-shift in clump G003.254+00.410 for HCO$^+$(1-0) and HCN(1-0) can be attributed to the possibility that G003.254+00.410 is located near the Galaxy center, where the UV radiation is strong. The centroid velocity differences between HCO$^+$(1-0) and N$_2$H$^+$(1-0) for other clumps may be caused by star formation activities, such as infall motions (see section 4.5). However, the reason why the HCN(1-0) in some clumps greatly shift (red or blue) from the systemic velocities is still unclear. Maybe more observations are needed to verify this result.

\subsection{Column densities and abundances of the molecules}
\begin{figure}
  \centering
\includegraphics[angle=0,scale=0.60]{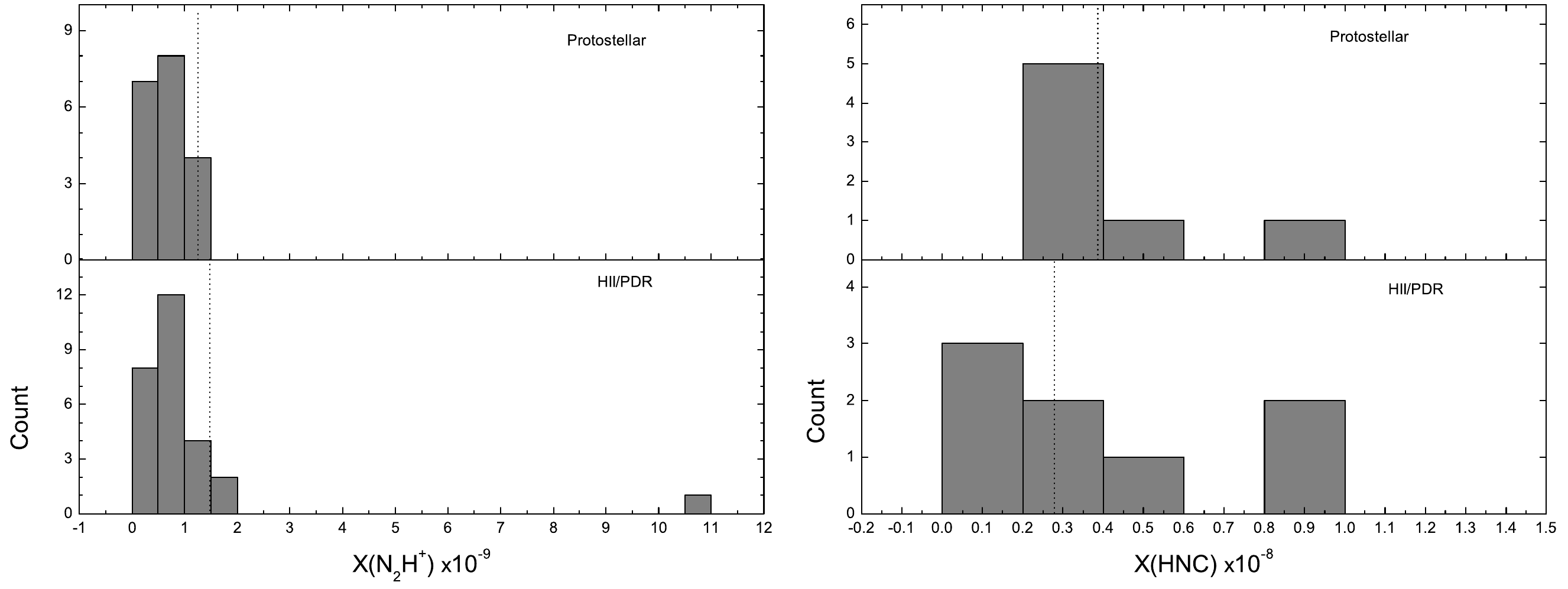}
\includegraphics[angle=0,scale=0.60]{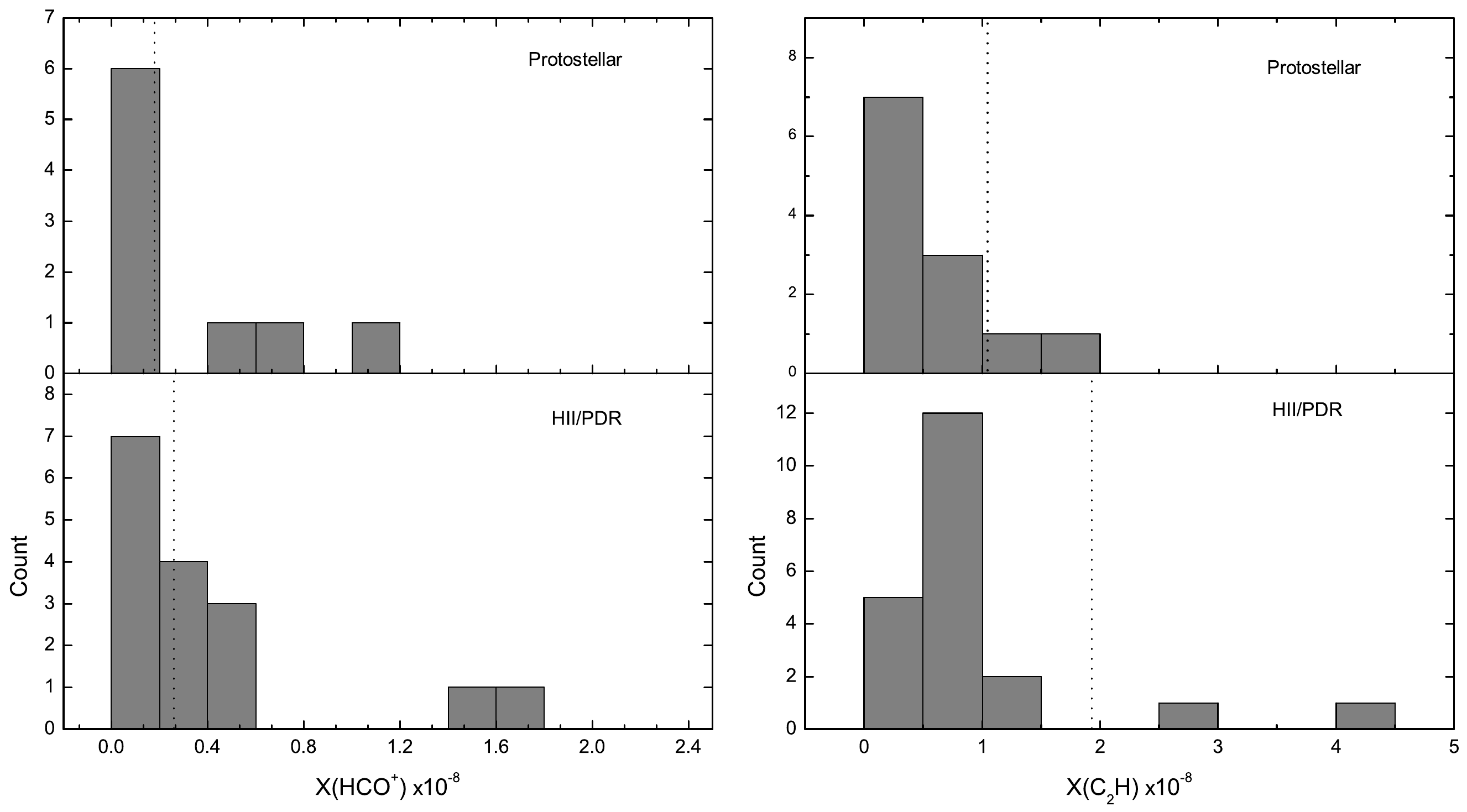}
\includegraphics[angle=0,scale=0.32]{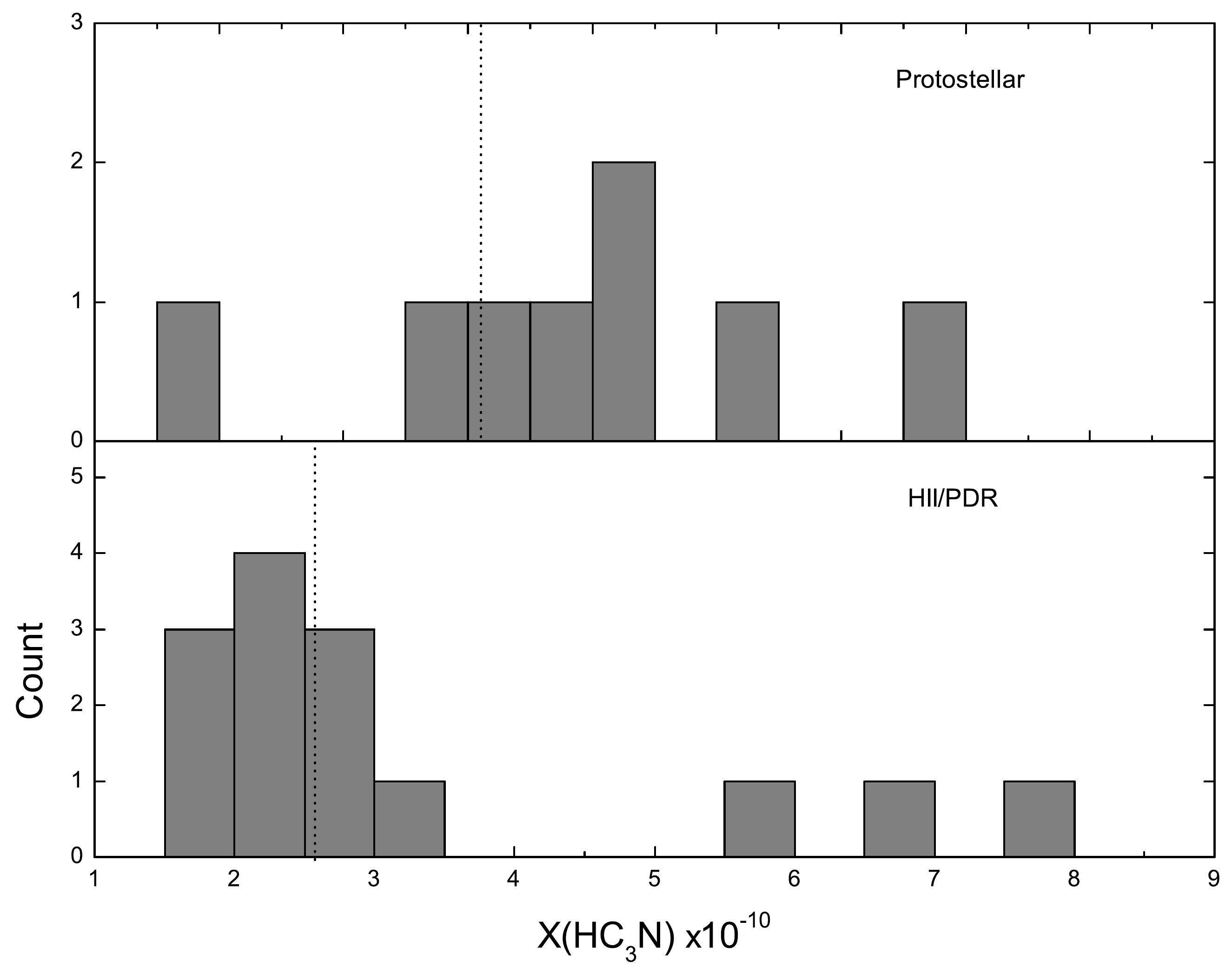}
\caption{Abundances histograms of N$_2$H$^+$, HNC, HCO$^+$, C$_2$H and HC$_3$N in protostellar clumps and HII/PDR regions, respectively. The dashed vertical lines indicate their median values.}
\label{fig9}
\end{figure}

Under the assumption of local thermodynamic equilibrium (LTE) and a beam filling factor of 1, the column density of the molecular lines can be derived from \citep{Sanhueza12}
\begin{equation}
\rm  N = \frac{3k^2}{16\pi^3\mu^2 h B^2 R_{in}}\frac{T_{ex}+hB/3k}{(J+1)^2}\frac{e^{E_u/k T_{ex}}}{e^{h\nu/k T_{ex}}-1}\times \frac{1}{J(T_{ex})-J(T_{bg})} \frac{\tau}{1-e^{-\tau}}\int{T_{MB}d\upsilon},
\label{eq8}
\end{equation}
where k is the Boltzmann constant, h is the Planck constant, $\mu$ is the permanent dipole moment of the molecule, $\nu$ is the transition frequency, J is the rotational quantum number of the lower state, E$\rm_u$ is the energy of the upper level, B is the rotational constant of the molecule, T$\rm_{bg}=2.73$ is the background temperature, T$\rm_{ex}$ is the excitation temperature. Here we assume that T$\rm_{ex}= T_{dust}$ and J(T) is defined by $\rm J(T)=(e^{h\nu/kT}-1)^{-1}$. $\rm \int{T_{MB}d\upsilon}$ is the integrated intensity. The values of $\mu$, B, and E$\rm_u$ can be obtained from the Cologne Database for Molecular Spectroscopy \footnote{http://www.astro.uni-koeln.de/cdms/} \citep[CMDS;][]{Muller01,Muller05}. And $\rm R_{in}$ is the relative intensity of the brightest hyperfine transition takeing into account the satellite lines corrected by their opacities. It is 5/9 for N$_2$H$^+$, 5/12 for C$_2$H, and fixed to 1.0 for transitions without hyperfine structure (hfs).

As for the calculation of the optical depth, assuming the same filling factor and excitation temperature, it can be derived from the intensity ratio of two isotopes spectra, one of which is optically thin.
\begin{equation}
\rm \frac{1-e^{-r\tau}}{1-e^{-\tau}}\simeq\frac{T_{MB,1}}{T_{MB,2}}
\label{eq9}
\end{equation}
where r is the intensity ratio of line1 over line2, and line2 is optically thin. $\rm T_{MB,1}$ and $\rm T_{MB,2}$ are the main beam brightness temperatures obtained from the gaussian fittings. Therefore, both the optical depths of HN$^{13}$C(1-0) and H$^{13}$CO$^+$(1-0) can be estimated through the isotope abundance ratios of $\rm  r\sim [^{12}C]/[^{13}C]$. In our work, we assume $\rm[^{12}C]/[^{13}C]\simeq50$ \citep{Purcell06}. The calculated results of the optical thickness of HN$^{13}$C(1-0) and H$^{13}$CO$^+$(1-0) are listed in columns 2 and 3 of Table~\ref{tab4}.

Actually, the molecular lines exhibiting hfs can also derive their optical depths via equation~(\ref{eq9}) by using the intensity ratio of two components \citep{Yu16}.
For C$_2$H(1-0), two components $\rm N_{J,F} = 1_{3/2,2}-0_{1/2,1}$ and $\rm N_{J,F} = 1_{3/2,1}-0_{1/2,0}$ are detected in our work (see the spectra in Fig.~\ref{fig7}) and the intensity ratio $\rm \frac{C_2H (F=2-1)}{C_2H (F=1-0)}$ of them is 2.0 \citep{Tucker74}.
But for N$_2$H$^+$(1-0), we substitute $\rm \int{T_{MB}d\upsilon}$ for $\rm T_{MB}$. Because the seven hfs are mixed into three groups with an intensity ratio of $1:5:3$ \citep{Yu16}. Consequently, the optical thickness for N$_2$H$^+$(1-0) can be obtained from following equation
\begin{equation}
\rm \frac{1-e^{-r\tau}}{1-e^{-\tau}}\simeq\frac{\int{T_{MB,1}d\upsilon}}{\int{T_{MB,2}}d\upsilon}
\label{eq10}
\end{equation}
Here $\rm \int{T_{MB,1}d\upsilon}$ is the second component N$_2$H$^+$ ($\rm N_{J,F} = 1_{2,3}-0_{1,2}$) and $\rm \int{T_{MB,2}d\upsilon}$ is the third component N$_2$H$^+$ ($\rm N_{J,F} = 1_{1,2}-0_{1,2}$). Therefore, the intensity ratio r here is 5/3. The derived optical depths of N$_2$H$^+$(1-0) and C$_2$H(1-0) are listed in columns 4 and 5 of Table~\ref{tab4}.

A lot of researches suggest HC$_3$N(10-9) to be optically thin \citep[e.g.][]{Chen13,Yu16}. In order to compare our work with other researches, such as \citet{Vasyunina11} and \citet{Sanhueza12}, here we also assume that HC$_3$N(10-9) is optically thin. Then through equation~(\ref{eq8}), we get the column densities of N$_2$H$^+$(1-0), HN$^{13}$C(1-0), H$^{13}$CO$^+$(1-0), C$_2$H(1-0) and HC$_3$N(10-9). Adopting the abundance ratio of $\rm [^{12}C]/[^{13}C] \sim 50$, the column densities of HNC(1-0) and HCO$^+$(1-0) are available. The derived results are showed in Table~\ref{tab4}. We also calculate their abundances through the formula of $\rm X_{line}=N_{line}/N_{H_2}$. Table~\ref{tab5} includes the calculated results. From Tables~\ref{tab4} and ~\ref{tab5}, we find that the median column densities of N$_2$H$^+$(1-0), HNC(1-0), HCO$^+$(1-0), C$_2$H(1-0) and HC$_3$N(10-9) are $(4.05\pm0.18)\times10^{13}$, $(1.40\pm0.28)\times10^{14}$, $(0.89\pm0.20)\times10^{14}$, $(4.90\pm0.74)\times10^{14}$, and $(1.22\pm0.12)\times10^{13}$ cm$^{-2}$ corresponding to the median abundances of $(1.42\pm0.13)\times10^{-9}$, $(0.28\pm0.07)\times10^{-8}$, $(0.25\pm0.05)\times10^{-8}$, $(1.85\pm0.26)\times10^{-8}$, and $(2.58\pm0.50)\times10^{-10}$ in HII/PDR regions, and $(2.92\pm0.16)\times10^{13}$, $(1.2\pm0.2)\times10^{14}$, $(0.69\pm0.16)\times10^{14}$, $(3.67\pm0.58)\times10^{14}$, and $(1.28\pm0.17)\times10^{13}$ cm$^{-2}$ corresponding to the median abundances of $(1.21\pm0.13)\times10^{-9}$, $(0.34\pm0.08)\times10^{-8}$, $(0.17\pm0.07)\times10^{-8}$, $(1.10\pm0.22)\times10^{-8}$, and $(3.86\pm0.58)\times10^{-10}$ in the proto-stellar clumps. Compared with the previous studies \citep{Vasyunina11,Sanhueza12,Yu16}, the column densities of the five species are at the same magnitude with them, but the abundances of the species are at a magnitude lower than those of \citet{Vasyunina11} and \citet{Sanhueza12} while are consistent with those of \citet{Yu16}. It implies that the BGPS clumps are less denser than the infrared dark clouds (IRDCs) but similar to the Red Midcourse Space Experiment Sources (RMSs). In addition, we show the histograms of N$_2$H$^+$(1-0), HNC(1-0), HCO$^+$(1-0), C$_2$H(1-0) and HC$_3$N(10-9) abundances in proto-stellar and HII/PDR stages in Fig.~\ref{fig9}. From Fig.~\ref{fig9}, we find that the abundances of Nitrogen-bearing molecules seem to decrease with evolution, while those of Oxygen-bearing molecules and C$_2$H increase.

\begin{figure*}
\begin{minipage}[t]{0.5\linewidth}
  \centering
   \includegraphics[angle=0,scale=0.25]{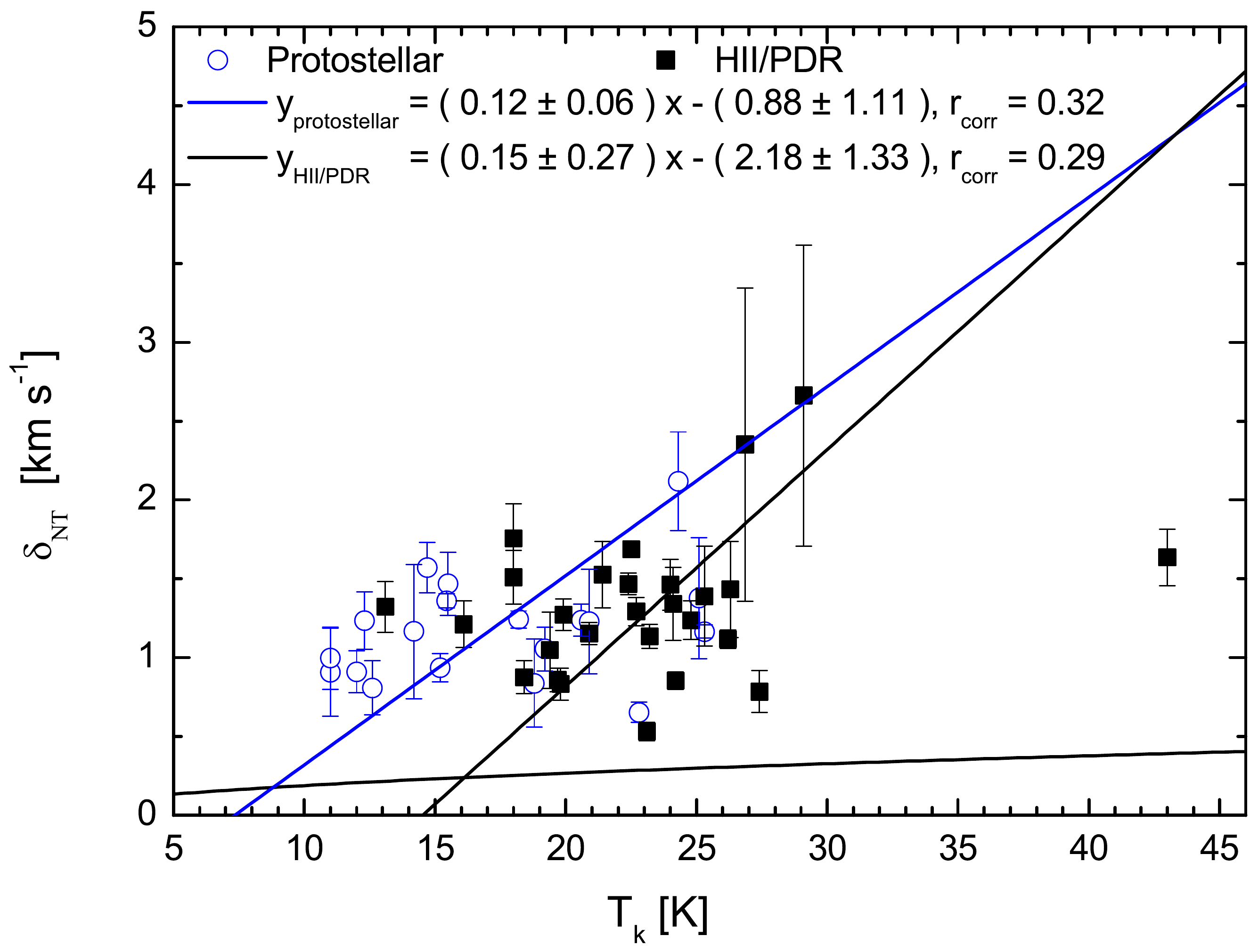}
  \end{minipage}%
  \begin{minipage}[t]{0.5\linewidth}
  \centering
   \includegraphics[angle=0,scale=0.25]{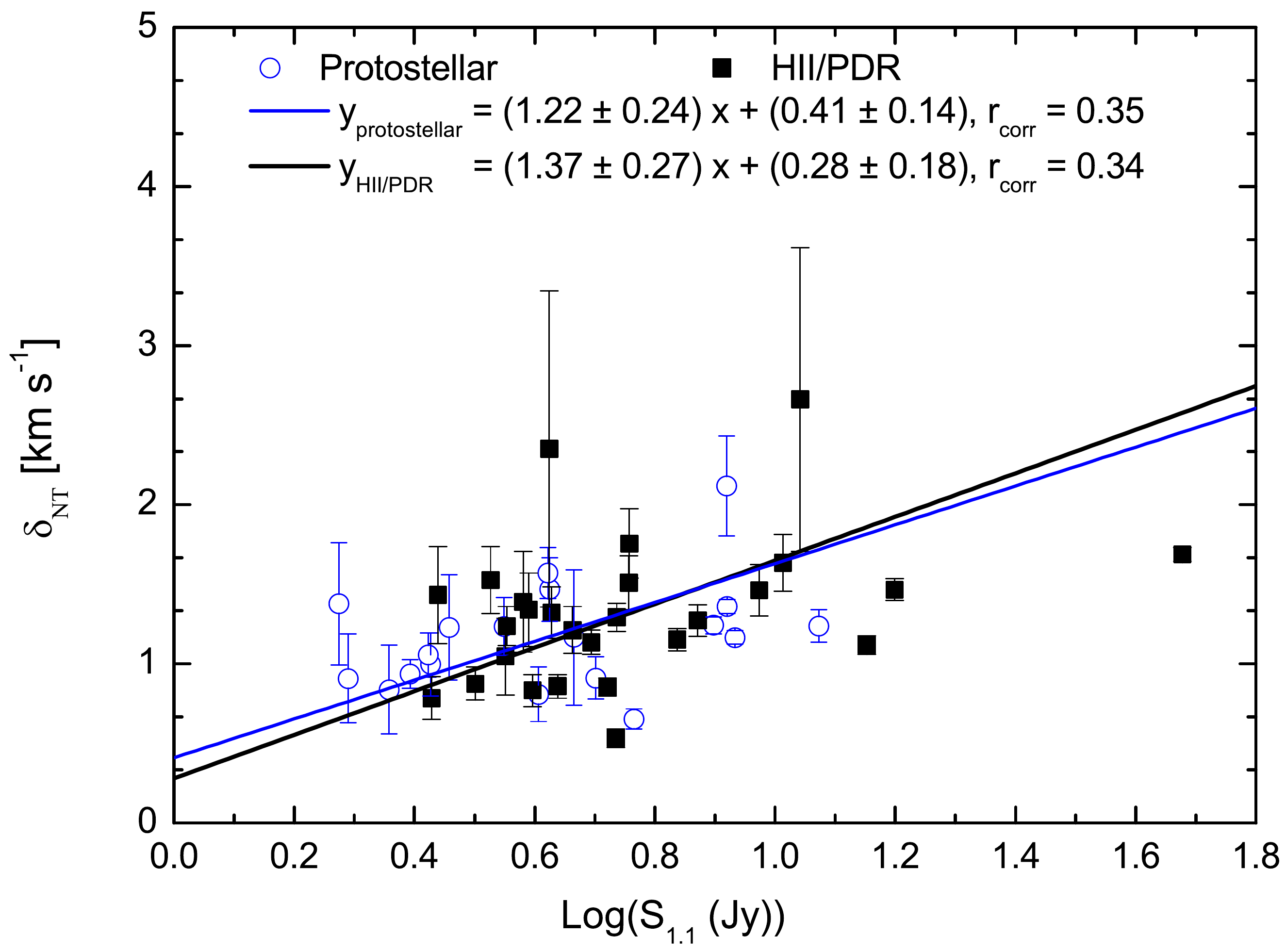}
  \end{minipage}%
\caption{Left panel: non-thermal velocity dispersion of N$_2$H$^+$ vs. the kinetic temperature. The black squares are for HII/PDR regions and the blue open circles represent the proto-stellar clumps. The black curve denotes the thermal sound speed given by $\rm a=(kT_{kin}/\mu m_H)^{1/2}$. Right panel: non-thermal velocity dispersion of N$_2$H$^+$ vs. the integrated 1.1 mm flux density. The black and blue solid lines are the least squares bisector fittings toward the HII/PDR regions and the proto-stellar clumps, respectively.}
\label{fig10}
\end{figure*}

\begin{figure}
  \centering
\includegraphics[angle=0,scale=0.60]{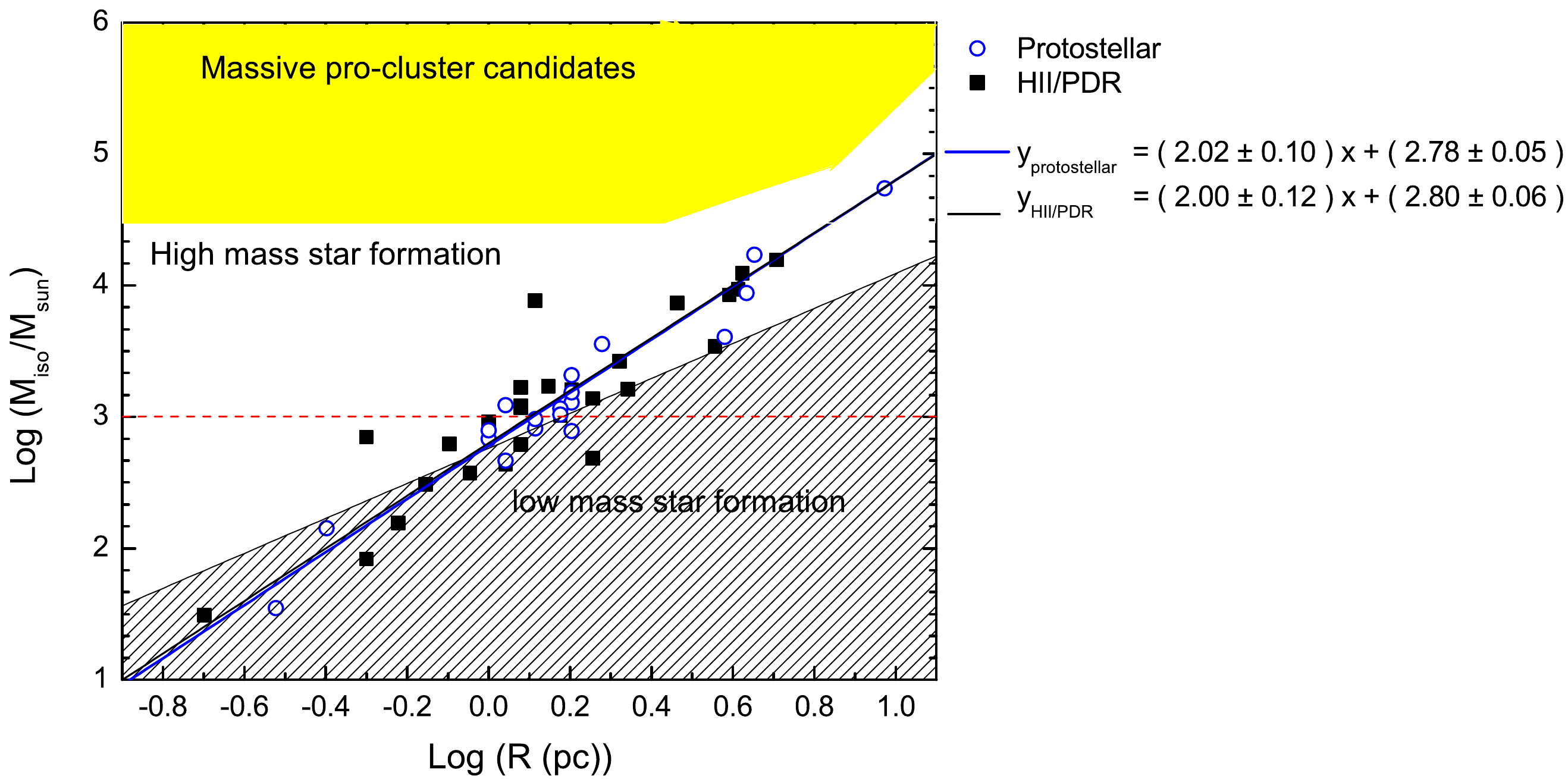}
\caption{The mass-size relationships of HII/PDR regions (black squares) and the proto-stellar clumps (blue open circles) with mass determined. The black shaded region represents the parameter space to be devoid of massive star formation, where $\rm Log(M/M_\odot)=1.33Log(R \,pc^{-1})+2.76$ \citep{Kauffmann10}. The yellow shadowed area indicates the region in which young massive cluster progenitors are expected to found \citep[e.g.][]{Bressert12}. The solid black and blue lines show the least squares bisector fittings to HII/PDR regions and proto-stellar clumps, separately. The red dashed horizon line represents $\rm M=1000\,M_\odot$.}
\label{fig11}

\end{figure}

\section{Discussion}
\subsection{Non-thermal velocity}
Since N$_2$H$^+$(1-0) is optically thin and detected in all the clumps, this molecule can cast light upon the internal properties of clumps. The line width of N$_2$H$^+$(1-0) responses to the internal motions within clumps, such as thermal motions, turbulence, inflows and outflows. In Table~\ref{tab3}, G003.254+00.410, which has the highest velocity dispersion of 11.1 km s$^{-1}$ will not be discussed in this subsection because of the complexity near the Galactic center. The velocity dispersions of our sample are in the range of $1.27-6.27$ km s$^{-1}$, with up to $ 82\%$ of clumps having $\rm \sigma_{obs}\geq 2.0 \, km \,s^{-1}$. By contrast to the previous studies \citep{Anglada96, Dunham11} with a range of $1-4$ km s$^{-1}$ and 98$\%$ of their sources less than $\leq2$ km s$^{-1}$, most of our sources are more active internally.

The observed line width is attributed to thermal and non-thermal motions of the gas. We can determine the non-thermal velocity dispersion by removing the thermal contribution to the observed line width through following express:
\begin{equation}
\rm \sigma_{NT} = \sqrt{\frac{\sigma_{obs}^2}{8ln2}-\frac{kT_{k}}{29m_H}},
\label{eq11}
\end{equation}
in which $\rm \sigma_{obs}$ is the observed N$_2$H$^+$(1-0) FWHM (in Table~\ref{tab3}). k is the Boltzmann constant and T$\rm_{k}$ is the kinetic temperature of the gas (here approximated to the dust temperature). Here $\rm \frac{kT_{kin}}{29m_H}$ is the thermal broadening. The median values of $\rm \sigma_{NT}$ in proto-stellar clumps and HII/PDR regions are $1.17\pm 0.17$ and $1.29\pm0.13$ km s$^{-1}$, respectively. Moreover, we plot the relations between $\rm \sigma_{NT}$ and $\rm T_{kin}$ as well as the integrated 1.1 mm flux density (S$_{1.1}$) in Fig.~\ref{fig10}. The blue open circles represent proto-stellar clumps and the black squares are HII/PDR regions. The black curve in left panel marks the thermal sound speed as $\rm a=(kT_{kin}/\mu m_H)^{1/2}$, which is at least three times smaller than the corresponding non-thermal line width. It indicates that the non-thermal motions dominate the line broadening of N$_2$H$^+$ in these clumps. And the low correlation coefficients around 0.3 of $\rm \sigma_{NT}$ with $\rm T_{kin}$ and S$_{1.1}$ separately suggest that the temperature and the 1.1 mm flux density seem to have an influence on the non-thermal motions to some extent. Furthermore, $\rm \sigma_{NT}$ as a whole probably increases with $\rm T_{kin}$ as well as Log(S$_{1.1}$) in each stage.

\begin{figure}
  \centering
\includegraphics[angle=0,scale=0.35]{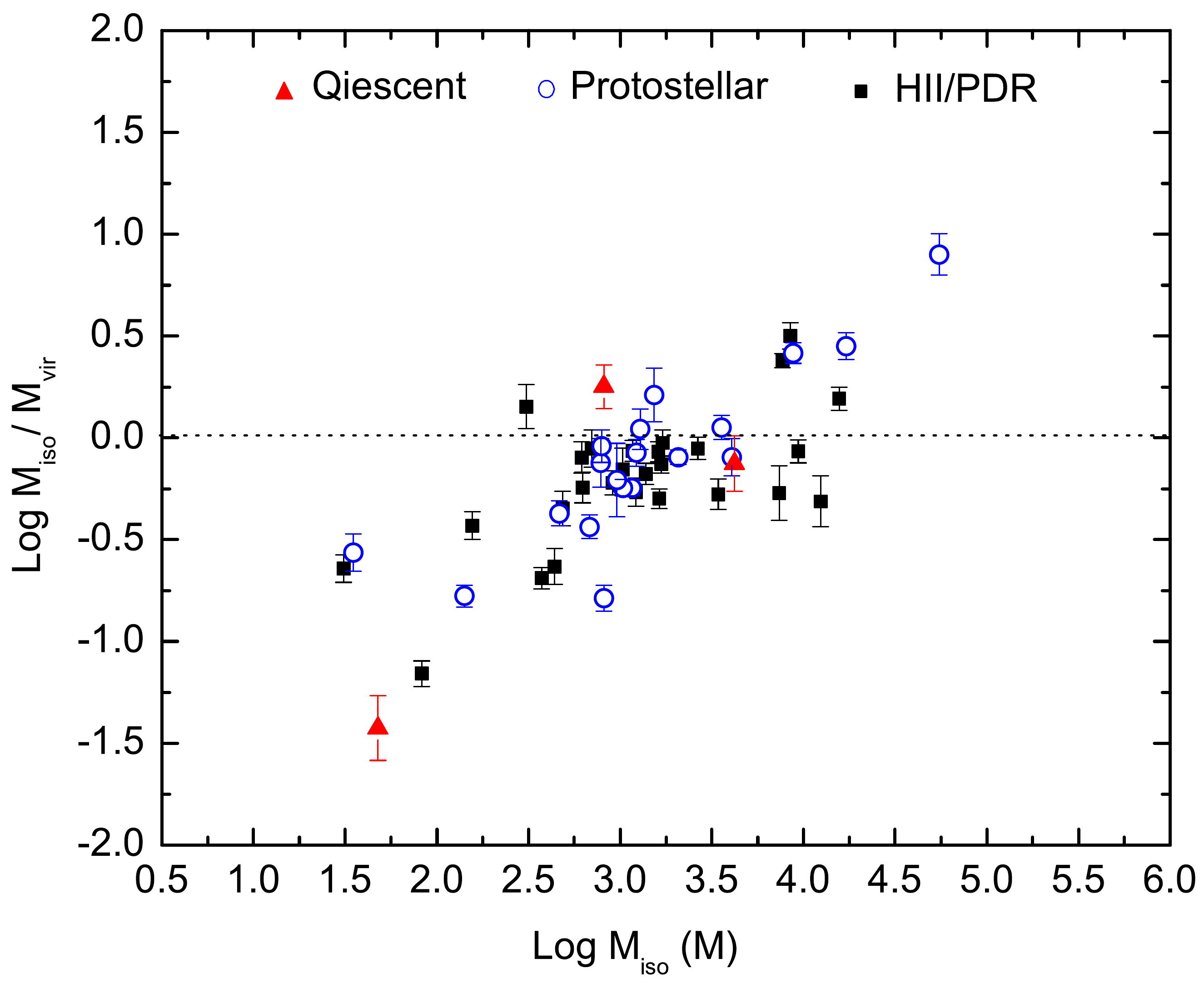}
\caption{Virial ratio $\rm Log \,(M_{iso}/M_{vir}$) as a function of $\rm Log\, (M_{iso}/M_\odot)$ for HII/PDR regions (black squares), proto-stellar clumps (blue open circles) and pre-stellar clumps (red triangles). The black horizon line indicates the locus of gravitational equilibrium for thermal and kinematic energies.}
\label{fig12}
\end{figure}

\subsection{The mass-size relationships}
Figure~\ref{fig11} presents the relationships between mass and size in proto-stellar and HII/PDR stages, respectively. The black shadowed region there represents a parameter space devoid of massive star formation, determined by \citet{Kauffmann10}. The yellow filled area indicates the region of massive pro-cluster candidates \citep[e.g.][]{Bressert12}. The red horizon dashed line marks M$\rm= 10^{3}\, M_\odot$. And, the solid blue and black lines separately indicate the least squares bisector fittings to them expressed as empirical relations of $\rm Log(M/M_\odot)_{protostellar}=(2.02\pm0.10)Log(R \,pc^{-1})+(2.78\pm0.05)$ with a correlation coefficient of 0.97, and $\rm Log(M/M_\odot)_{HII/PDR}=(2.00\pm0.12)Log(R \,pc^{-1})+(2.80\pm0.06)$ with a correlation coefficient of 0.90. The mass and radius show a close positive correlation and a power-law relationship with each other. When M$\geq 10^{3}\, M_\odot$, the clumps have a high potential to form high-mass stars, otherwise, the chance of a clump to form a massive star is merely up to 28$\%$. As a whole, based on Fig.~\ref{fig11}, about 74.1$\%$ HII/PDR regions and 78.9$\%$ proto-stellar clumps have an opportunity to form high-mass stars. Moreover, the slop index of the mass-radius in our work seems to be a constant around 2.00, which is similar to that of \citet{He16} considering the uncertainty. But more explorations are required to justify this conclusion.
\begin{figure}[t]
\begin{minipage}[t]{0.5\linewidth}
  \centering
   \includegraphics[angle=0,scale=0.25]{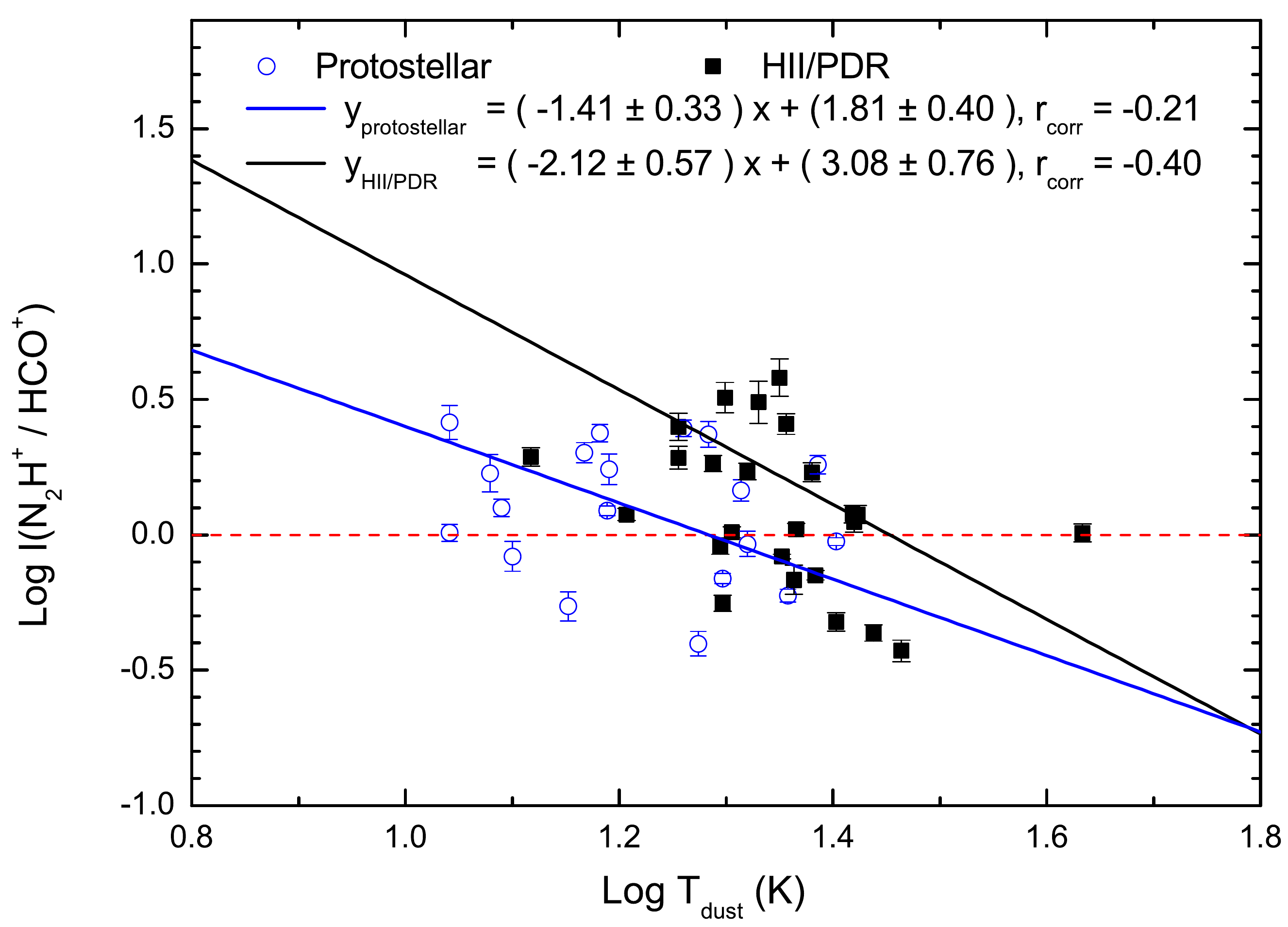}
  \end{minipage}%
  \begin{minipage}[t]{0.5\linewidth}
  \centering
   \includegraphics[angle=0,scale=0.25]{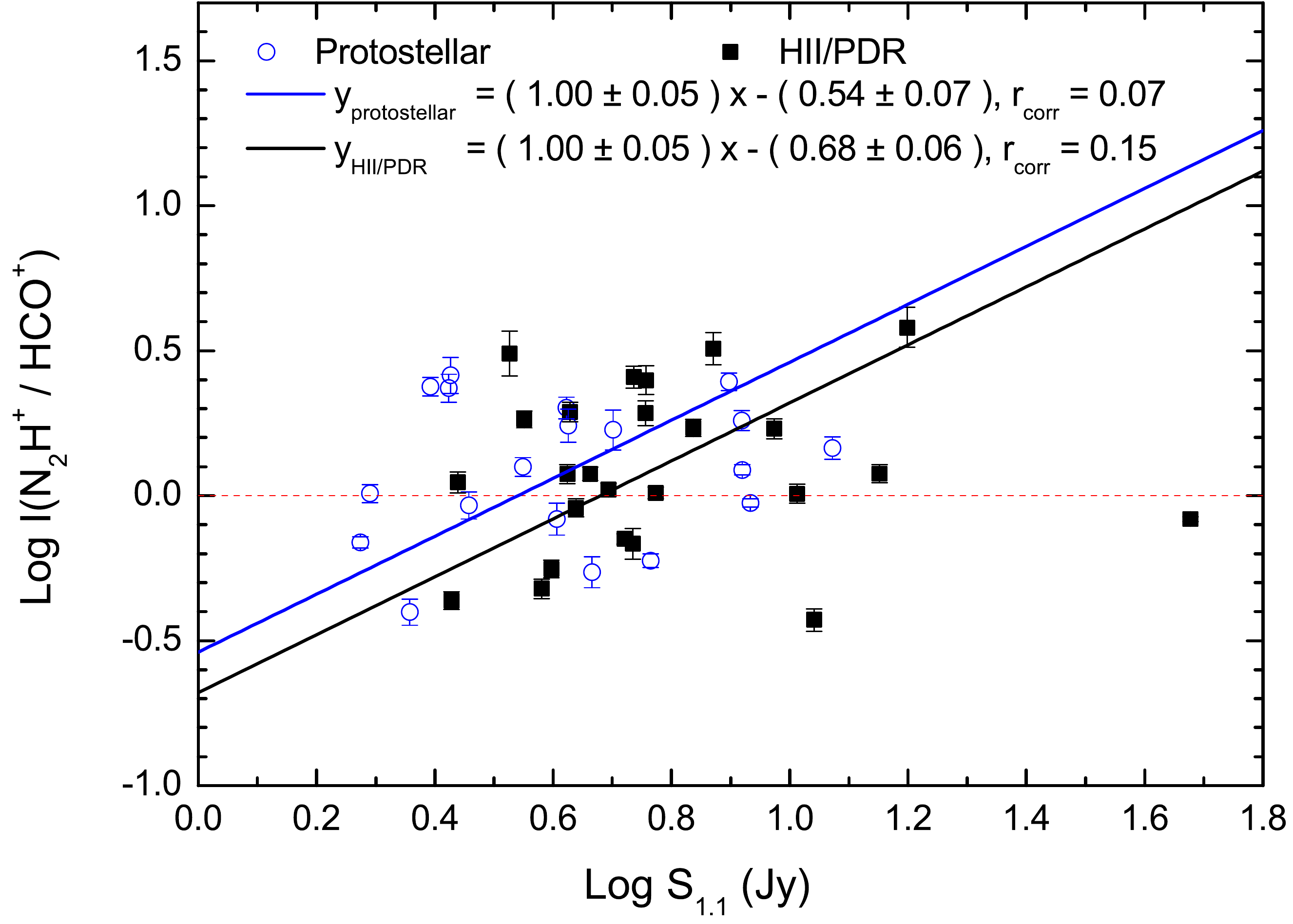}
  \end{minipage}%
  \vspace{2mm}
  \begin{minipage}[t]{0.5\linewidth}
  \centering
   \includegraphics[angle=0,scale=0.25]{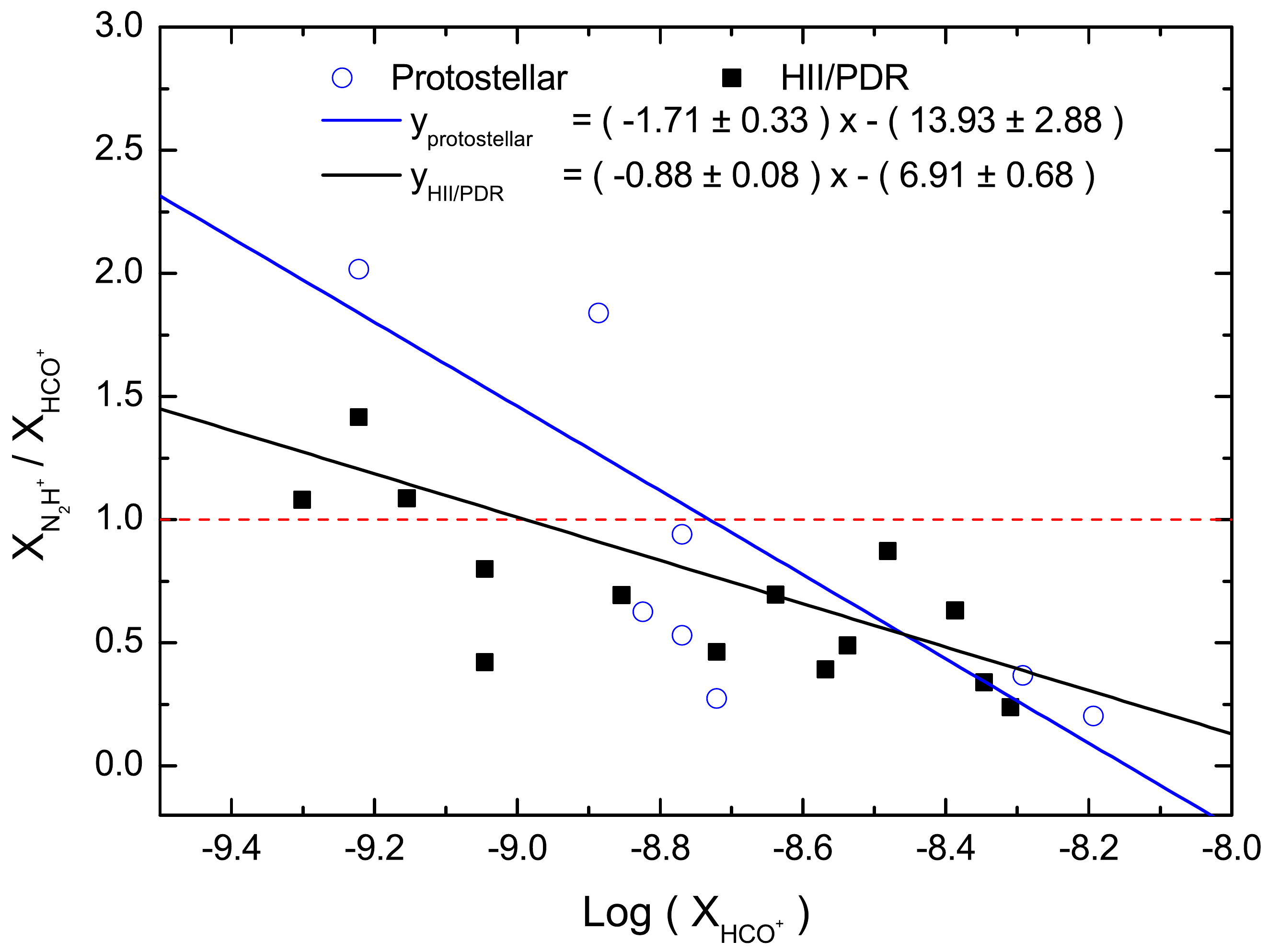}
  \end{minipage}%
  \begin{minipage}[t]{0.5\linewidth}
  \centering
   \includegraphics[angle=0,scale=0.25]{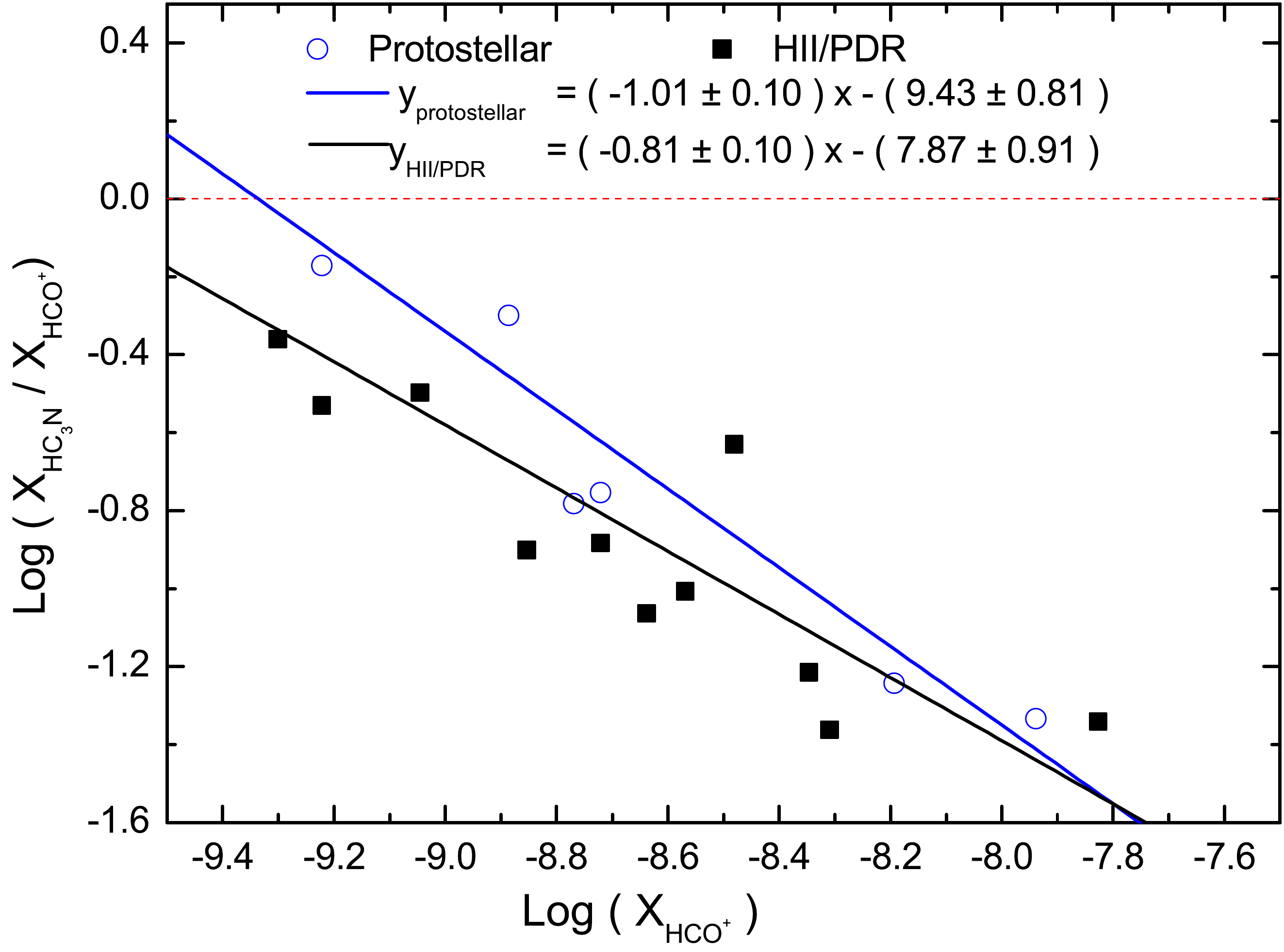}
  \end{minipage}%
\vspace{2mm}
  \begin{minipage}[t]{0.5\linewidth}
  \centering
   \includegraphics[angle=0,scale=0.25]{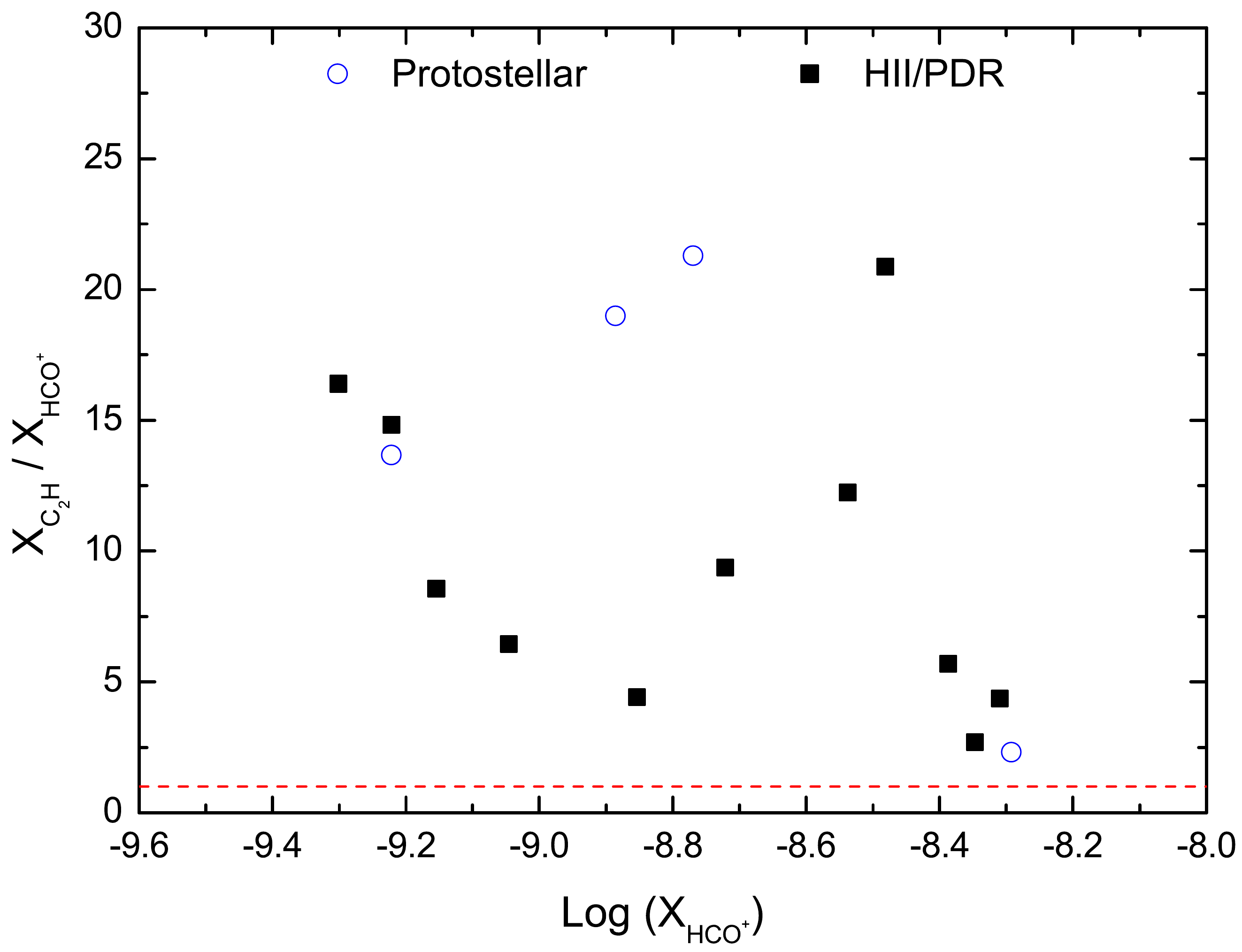}
  \end{minipage}%
  \begin{minipage}[t]{0.5\linewidth}
  \centering
   \includegraphics[angle=0,scale=0.25]{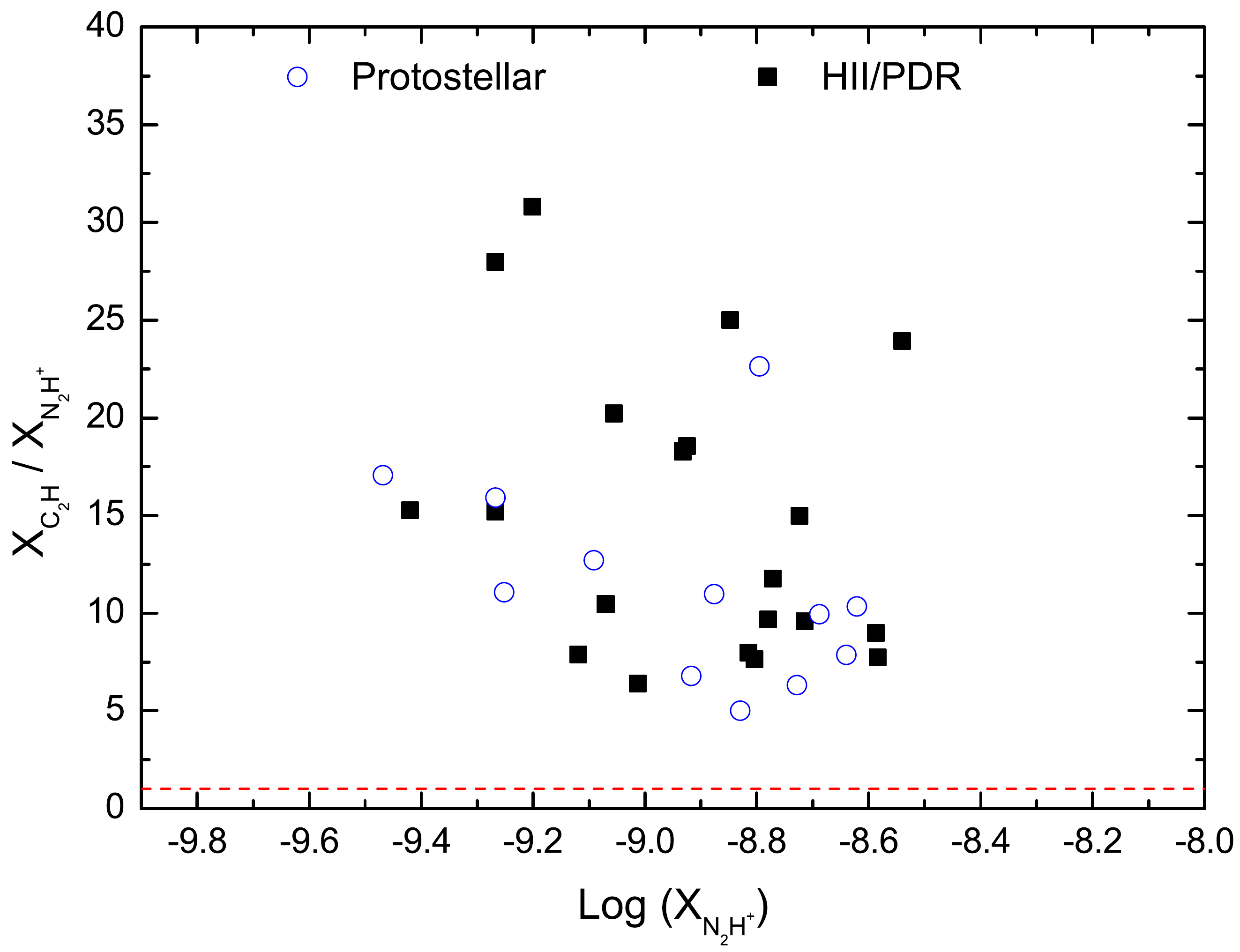}
  \end{minipage}%
  \caption{Top-left: plot integrated intensity ratio Log(N$_2$H$^+$/ HCO$^+$) vs. the dust temperature Log T$\rm {dust}$ (K). Top-right: plot integrated intensity ratio Log(N$_2$H$^+$/ HCO$^+$) vs. the integrated 1.1 mm flux Log S$_{1.1}$(Jy). Middle-left: plot abundance ratio X$\rm_{N_2H^+}$/X$\rm_{HCO^+}$ vs. Log (X$\rm_{HCO^+}$). The black and the blue line are the linear fittings for the clumps in two different evolutionary stages (HII/PDR and proto-stellar). Middle-right: plot abundance ratio Log(X$\rm_{HC_3N}$/X$\rm_{HCO^+}$) vs. Log (X$\rm_{HCO^+}$). The black and the blue curves indicate the power fittings at two different evolutionary stages-HII/PDR (black squares) and proto-stellar (blue open circles). Bottom-left: plot abundance ratio X$\rm_{C_2H}$/X$\rm_{HCO^+}$ vs. Log (X$\rm_{HCO^+}$) in proto-stellar clumps (blue open circles) and HII/PDR regions (black squres) separately. Bottom-right: plot abundance ratio X$\rm_{C_2H}$/X$\rm_{N_2H^+}$ vs. Log (X$\rm_{N_2H^+}$) in proto-stellar clumps (blue open circles) and HII/PDR regions (black squres) respectively. The red dashed lines in the subfigures represent y=1.}
  \label{fig13}
\end{figure}
\subsection{Virial mass}
Based on the analysis of \citet{He16}, we can assess the virial mass of each clump by utilizing the line width of N$_2$H$^+$(1-0) \citep{Urquhart13a}
\begin{equation}
\rm \frac{M_{vir}}{M_\odot} = \frac{783}{7ln2}(\frac{R}{pc})(\frac{\sigma_{avg}}{km\,s^{-1}})^2,
\label{eq12}
\end{equation}
where R is the effective radius that derived from equation~(\ref{eq6}) in Table~\ref{tab1} and $\rm \sigma_{avg}$ is the average velocity dispersion of gas via
\begin{equation}
\rm \sigma_{avg} = (\sigma_{obs})^2+8ln2\times \frac{kT_{kin}}{m_H}(\frac{1}{\mu_p}-\frac{1}{\mu_{N_2H^+}})
\label{eq13}
\end{equation}
where $\rm \mu_p\simeq2.33$ and $\rm \mu_{N_2H^+}=29$ are the mean molecular masses of molecular hydrogen and N$_2$H$^+$, respectively. Since N$_2$H$^+$(1-0) lines were optically thin in all the clumps (see the $\tau$ values of N$_2$H$^+$(1-0) in Table~\ref{tab4}) and the spectra were extracted from the peak positions, $\rm \sigma_{avg}$ are overestimated which results in the overestimated $\rm M_{vir}$. The virial masses of 50 sources are presented in Table~\ref{tab1}. The uncertainties in the parentheses come from measurement errors of the line widths.

We plot the virial ratios $\rm Log(M_{iso}/M_{vir})$ versus the clump masses in the stages from pre-stellar to proto-stellar to HII/PDR in Fig.~\ref{fig12}. The dashed horizon line marks the line of gravitational stability. Sources with ratios $\rm Log(M_{iso}/M_{vir})$ over this line are gravitationally bound and then collapse to possibly form stars, otherwise the clumps may be at a state of gravitationally unbound. In Fig.~\ref{fig12}, we find that over 65$\%$ sources in various evolutionary stages are gravitationally unbound and have no opportunity to form stars at present, which is paradoxical to the indicators of star formation, such as SiO (2-1) emission, infall motions (discussed in section 4.5) and 24 $\mu$m sources. The likely explanation is that most of the clumps contain a set of sub-cores rather than only one core and hence most of the clumps' envelopes are gravitationally unbound.

\begin{figure}
\begin{minipage}[t]{0.5\linewidth}
  \centering
   \includegraphics[angle=0,scale=0.5]{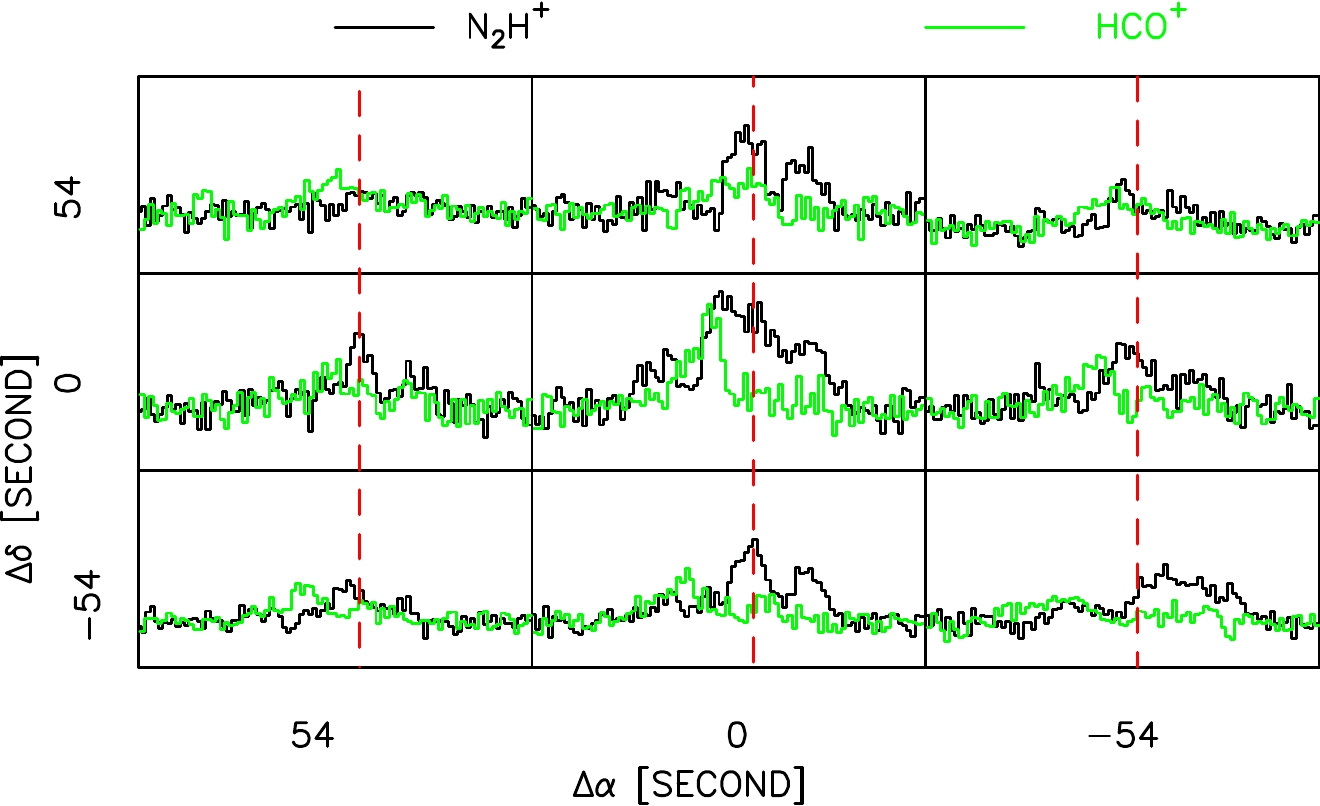}
  \end{minipage}%
  \begin{minipage}[t]{0.5\linewidth}
  \centering
   \includegraphics[angle=0,scale=0.5]{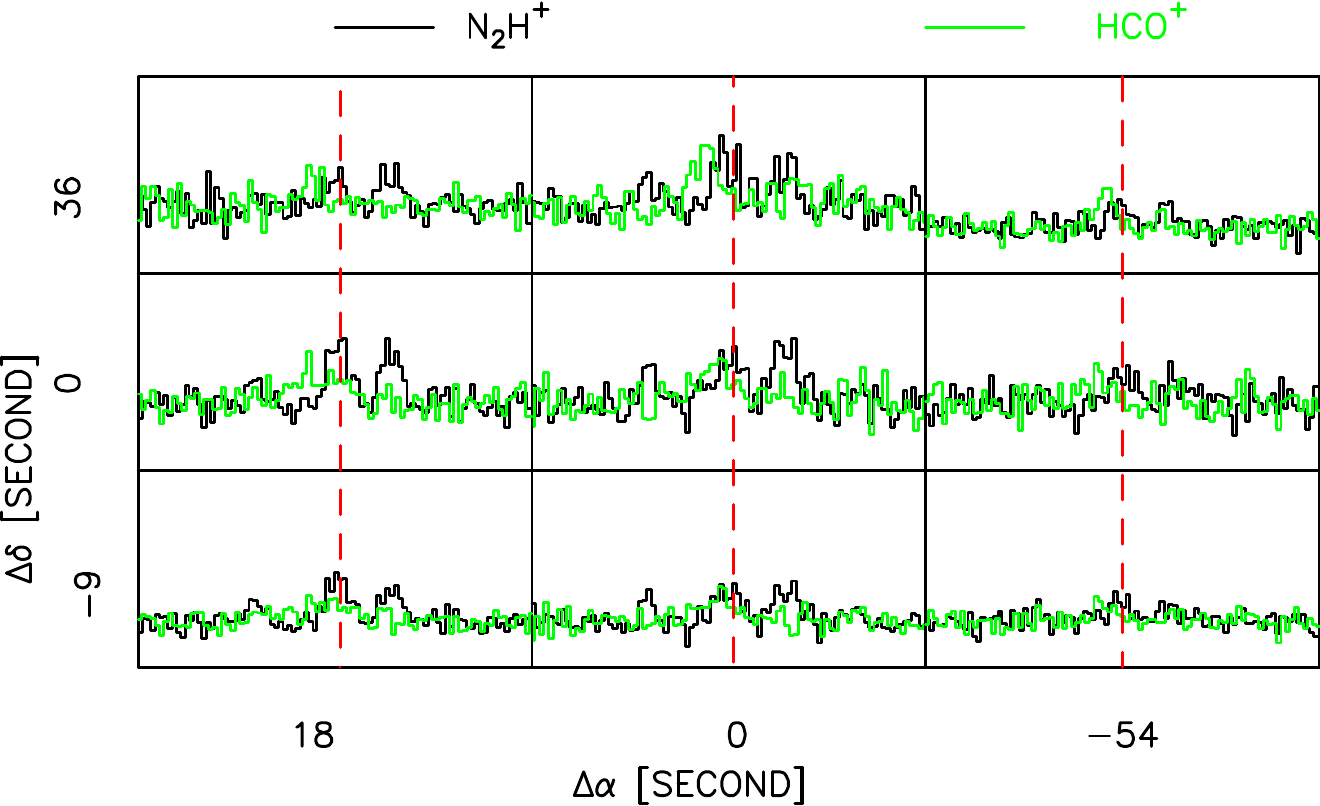}
  \end{minipage}%
  \vspace{2mm}
  \begin{minipage}[t]{0.5\linewidth}
  \centering
   \includegraphics[angle=0,scale=0.5]{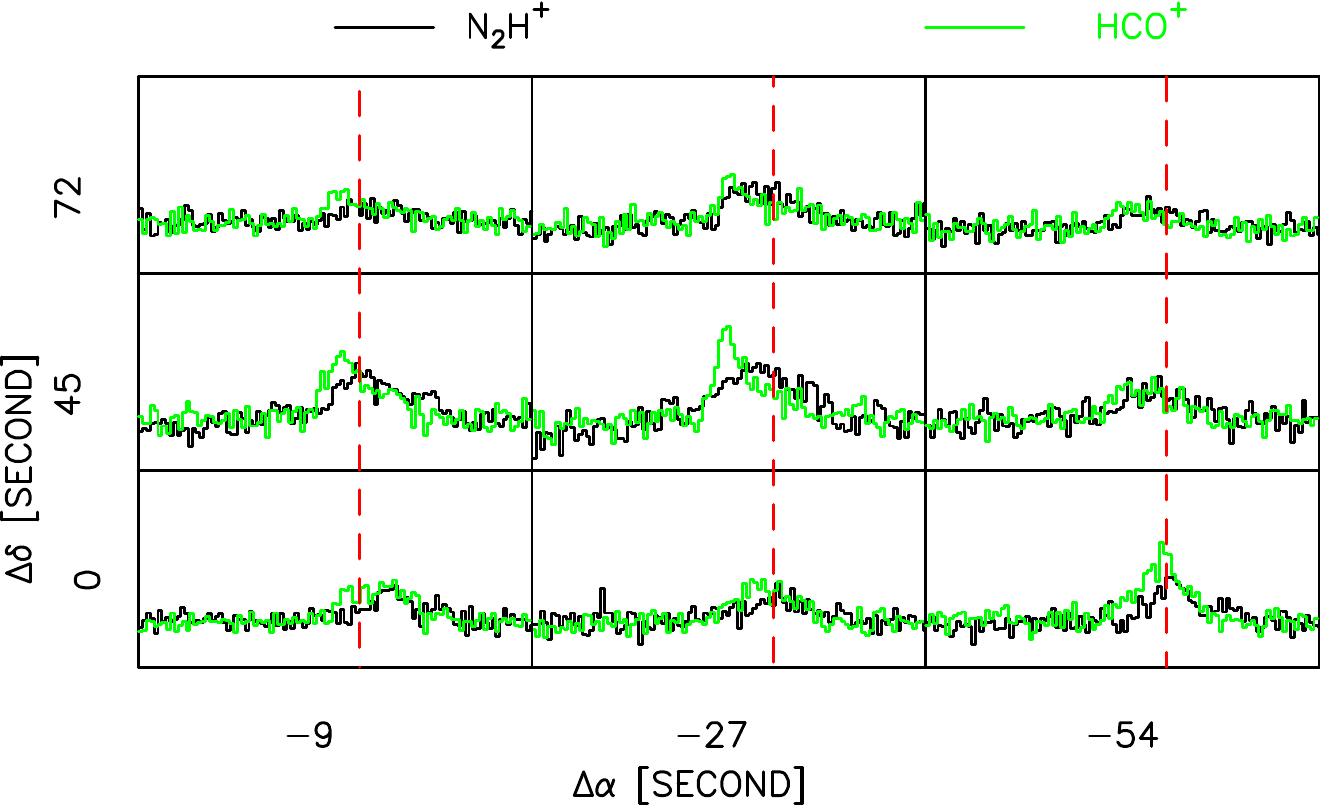}
  \end{minipage}%

  \caption{The mapping grids for the three infall candidates G010.214-00.324 (top-left), G011.121-00.128 (top-right), and G012.215-00.118(a)(bottom-left). The black lines are for N$_2$H$^+$ and the green lines represent HCO$^+$. The grids also denote the scales where the infall motions exist.}
  \label{fig14}
\end{figure}

\subsection{Chemical correlations between different species}
The ratio of $\rm N_2H^+ /HCO^+$ emission can be a chemical indicator of the evolution of the clump, since N$_2$H$^+$ and HCO$^+$ have opposite chemical behavior with respect to CO. HCO$^+$ can be produced by two chemical reactions of $\rm H{_3}^+ +CO\rightarrow HCO^+ + H_2$ and $\rm N_2H^+ + CO\rightarrow HCO^+ +N_2$ \citep{Jorgensen04,Schlingman11}. As CO is freezed onto dust grains at high densities ($\rm n> 10^4 \, cm^{-3}$) and low temperatures (T$\rm_k <$ 20 K), the large ratio of $\rm N_2H^+ /HCO^+$ emission illustrates a cold and dense environment. Otherwise, with the temperature rising, CO is eventually free from the dust grains and then results in the destruction of N$_2$H$^+$ and production of HCO$^+$. Consequently, the integrated intensity ratio of $\rm N_2H^+ /HCO^+$ should decrease with the temperature. Besides, N$_2$H$^+$ does not deplete onto dust grains until densities reach $10^6$ cm$^{-3}$ \citep{Tafalla04,Flower06}, which can be excluded in our work (see Table~\ref{tab1}).

Figure~\ref{fig13} shows the variation of the integrated intensity ratio N$_2$H$^+$(1-0) to HCO$^+$(1-0) versus the dust temperature and the integrated flux at 1.1 mm towards two evolutionary stages (proto-stellar and HII/PDR). The integrated intensity ratios $\rm I(N_2H^+ / HCO^+)$ actually decline as power-law functions and have negative correlations ($\rm r_{corr-protostellar}=-0.21$ and $\rm r_{corr-HII/PDR}=-0.40$) with T$\rm_{dust}$ in both stages, especially in the later evolutionary stage. However, the relations between the ratio of $\rm N_2H^+ /HCO^+$ emission and the integrated flux are positive in these two stages despite of the lower correlation coefficients ($\rm r_{corr}(protostellar)=0.07$ and $\rm r_{corr}(HII/PDR)=0.15$), which means almost no correlations. Even so, we still make the least squares bisector fittings between Log $\rm I(N_2H^+ / HCO^+)$ and Log ($\rm S_{1.1}$) (see the top-right panel in Fig.~\ref{fig13}) and find two similar linear functions with both slops as 1.0. And the poor correlations further suggest that the BGPS clumps are unlikely to be single massive cores but probably consist of multiple smaller cores, between which the fraction of cold, dense and CO-depleted gas is different \citep{Shirley13}. Comparing these two subfigures in Fig.~\ref{fig13} (slops and correlation coefficients), we can find that under the density of $10^6$ cm$^{-3}$, temperature plays a more vital role in the chemical evolution of molecules.

Furthermore, we also present the abundance relations between four species (N$_2$H$^+$, HCO$^+$, C$_2$H and HC$_3$N) in Fig.~\ref{fig13} and find that the abundances of Nitrogen-bearing molecules (N$_2$H$^+$ and HC$_3$N) dramatically descend with those of the Oxygen-bearing molecule (HCO$^+$) in both proto-stellar and HII/PDR clumps. This probably suggests that HCO$^+$ produced by the destruction of the Nitrogen-bearing molecules and N$_2$H$^+$ is the primary contribution, since the abundance of HC$_3$N seems not to vary in the HCO$^+$ relatively more abundant region. On the other hand, the abundance of C$_2$H appears to fluctuate with HCO$^+$. It might be related to the major destruction route that C$_2$H is reacted with O to form CO \citep{Pan17}. And C$_2$H seems to have nothing to do with N$_2$H$^+$ according to their abundance relationship in Fig.~\ref{fig13}. Combined with Table~\ref{tab5}, the Nitrogen-bearing molecules are likely to be more abundant in the earlier stages while the Oxygen-bearing species and C$_2$H are adverse. Furthermore, C$_2$H is the most abundant species, followed by the sequence of HCO$^+$, N$_2$H$^+$, HNC, and HC$_3$N.

\subsection{Infall candidates}
Based on two methods to identify the infall candidates \citep[e.g.][]{Mardones97,Fuller05,Wu05,Sun09,Chen13,Liu14}, we find three new infall candidates G010.214-00.324, G011.121-00.128, and G012.215-00.118(a). In a quantization way, $\rm \delta V$ defined as $\rm \delta V=(V_{thick}-V_{thin})/\Delta V_{thin}$ is the asymmetry parameter to judge the infall motions with $\rm\delta V>0.25$ for red asymmetry and $\rm\delta V<-0.25$ for blue asymmetry. Here $\rm V_{thick}$ represents the line peak velocity of the optically thick line and $\rm V_{thin}$, an optically thin tracer, measures the systemic velocity. $\rm\Delta V_{thin}$ is the line width of the optically thin line. In our work, N$_2$H$^+$ represents the optically thin while HCO$^+$ and HNC are optically thick. The asymmetry parameters $\rm \delta V$ of HCO$^+$ and HNC for the three new discovered infall candidates are listed in Table~\ref{tab6} according to the observed parameters in Table~\ref{tab3}. Both G010.214-00.324 and G011.121-00.128 exhibit the blue asymmetry in HCO$^+$ and HNC lines, while G012.215-00.118(a) shows a blue asymmetry in HCO$^+$ line but non asymmetry in HNC line. Following the criterion utilized in \citet{He15}, infall candidates must have a blue skewed profile at least one optically thick line (HCO$^+$(1-0) or HNC(1-0)), no red skewed profile in the other optically thick line and no spatial difference in the mapping observation. These three clumps are likely to be infall candidates.

In order to assure that the infall motions really exist in large scales, we make the mapping grids for the three infall candidates in Fig.~\ref{fig14}. Each mapping shows the whole region where the blue shewed profiles occur, and hence works out the size of the infall motion. From Fig.~\ref{fig14}, we can obtain the infall motion sizes for G010.214-00.324, G011.121-00.128, and G012.215-00.118(a) are $\rm1.64\,pc\times1.64\,pc$ ($1.8'\times1.8'$), $\rm1.06\,pc\times0.66\,pc$ ($1.2'\times0.75'$) and $\rm3.05\,pc\times4.88\,pc$ ( $0.75'\times1.2'$). Therefore, these three clumps actually show the large-scaled infall motions. Furthermore, we roughly estimate their infall rate by the following formula \citep{Lopez-Sepulcre10}:
\begin{equation}
\rm \dot{M}_{inf} = 4\pi\mu_{H_2}m_H r^2 V_{inf} n=3.15\times10^{-4}(\frac{r}{pc})^2(\frac{V_{inf}}{km\,s^{-1}})(\frac{n}{cm^{-3}}),
\label{eq14}
\end{equation}
where $\rm V_{inf} = V_{N_2H^+} - V_{HCO^+}$ is a rough estimate of the infall velocity, $\rm r=\sqrt{r_{maj} r_{min}}$ is the equivalent radius of the infall region, and n is the volume density described in section 3.2. Through equation~(\ref{eq14}), we get the infall rates for G010.214-00.324, G011.121-00.128, and G012.215-00.118(a) to be $(3.4\pm0.4)\times10^{-3}$, $(1.6\pm0.2)\times10^{-3}$, and $(6.5\pm2.5)\times10^{-3}$ M$_\odot$ yr$^{-1}$, which have the similar magnitude with the previous studies \citep{Lopez-Sepulcre10,He16}.

\section{Summary}
\label{sect:summary}
We perform an investigation towards 48 high-mass clumps selected on the basis of the BGPS and MALT90 survey data and aim at studying the fragmentation, classification, physical and chemical features, chemical evolution of dense gases and star formation activities. Our main results are summarized as follows£º

1. Combining the 1.1 mm continuum data with the molecules emission, we find that only two clumps (G012.215-00.118 and G354.000+00.474) of our sample show the sub-structures and all the remaining clumps are the single-core structure. It might be attributed to the similar spatial resolutions between the BGPS continuum data ($33''$) and the MALT90 survey ($36''$). Among final 50 clumps, 27 are HII/PDR, 19 are pro-stellar, 3 are pre-stellar while 1 is uncertain. These clumps concentrate on the molecular ring and the expanding arm.

2. Eleven species are detected in our work. They are N$_2$H$^+$(1-0), HNC(1-0), HCO$^+$(1-0), HCN(1-0), HN$^{13}$C(1-0), H$^{13}$CO$^+$(1-0), C$_2$H(1-0), HC$_3$N(10-9), SiO(2-1), HNCO($4_{4,0}-3_{0,3}$) and $^{13}$CS(2-1). Among them,  N$_2$H$^+$(1-0) and HNC(1-0) are detected in all the fields, the detection rates of other N-bearing molecular lines decline with evolution except for HC$_3$N(10-9), while those of O-bearing spectra show an inverse regulation. The reason why the detection rate of HC$_3$N(10-9) rises with evolution may be that this molecule is related to the C-chain generation and is an indicator of the hot cores. And the comparison of their centroid line velocities of each clump indicates that nine molecular lines can provide a clump the accurate systemic velocity except HCO$^+$ and HCN, whose V$\rm_{LSR}$ can deviate up to 12 km s$^{-1}$ from those of other molecular transition lines.

3. The temperature, mass and volume density increase as the clumps evolve, while the H$_2$ column density first shows an increase from pre-stellar to proto-stellar then almost remain steady at HII/PDR stage. However, the clumps on the whole accumulate material efficiently as they evolve. And both mass distributions of the clumps at proto-stellar and HII/PDR focus on a median value around 1000 M$_\odot$.

4. The masses and sizes of the clumps in proto-stellar and HII/PDR stages both show close correlations with each other and present two power-law functions with the indexes of $2.02\pm0.10$ and $2.00\pm0.12$, respectively, which are sharper than that of \citet{Kauffmann10} (1.33). The similar indexes in two stages maybe suggest that the mass-size relationship does not vary with the evolution. More researches should be done to justify this conclusion. And more than 70$\%$ clumps at these two stages should be capable of forming high-mass stars, but the envelopes of most clumps are gravitationally unbound at the moment, indicating most of our clumps probably consist of several cores within rather than a single structure.

5. The abundance ratios of N-bearing species to O-bearing species decrease sharply as the clumps evolve, indicating the trend of destruction of N-bearing species and production of O-bearing species in the later evolutional phrase. Therefore, the ratios of N-bearing species to O-bearing molecules can be the chemical indicators of evolution. In addition, C$_2$H is the most abundant molecule in all the stages and its chemical evolution may pose a slight impact on that of HCO$^+$.

6. Three infall candidates are identified. They are G010.214-00.324, G011.121-00.128, and G012.215-00.118(a), separately exhibiting large-scaled infall motions with the infall rates of $(3.4\pm0.4)\times10^{-3}$, $(1.6\pm0.2)\times10^{-3}$, and $(6.5\pm2.5)\times10^{-3}$ M$_\odot$ yr$^{-1}$.

\normalem
\begin{acknowledgements}
We are very grateful to the anonymous referee for his/her
helpful comments and suggestions. This research has made use of the data products from the {\it
Millimetre Astronomy Legacy Team 90 GHz (MALT90)\/} survey, and also
used NASA/IPAC Infrared Science Archive, which is operated by the
Jet Propulsion Laboratory, California Institute of Technology, under
contract with the National Aeronautics and Space Administration. This work is supported
by the National Natural Science Foundation of China (No. 11363004 and Grant
No. 11403042). This work is supported by the National Key Basic Research Program of China ( No.2015CB857100 ).

\end{acknowledgements}

\clearpage
\begin{landscape}
\setlength{\LTleft}{0pt} \setlength{\LTright}{0pt}
{\tiny
 \renewcommand{\footnoterule}{}
  \begin{longtable}{p{0.6cm}llccccccccccccccr}
   \caption{\label{tab1}The basic and derived clump parameters. }\\
      \hline
ID   &BGPS clumps$^{\rm a}$     &l    &b     &\multicolumn{2}{c}{distance}  &$\rm T_{d}$  & \multicolumn{2}{c}{R}      &$\rm S_{40}$   &$\rm S_{int}$ &$\rm N_{H_2}$ &$\rm n_{H_2}$  &$\rm M_{iso}$       &$\rm M_{vir}$   &classification          \\
     &                &deg  & deg &kpc &ref                      &K           & arcsec  &pc                & Jy            & Jy           & $\rm\times 10^{22}\,cm^{-2}$      &$\rm\times10^3\,cm^{-3}$       & $\rm M_{\odot}$      &$\rm M_{\odot}$   &                                 \\

(1)  & (2)                  &(3) &(4)                      &(5)         & (6)     &(7)            &(8)           &(9)           &(10)               &(11)           &(12)     &(13)     &(14) &(15) &(16)\\
\hline
\endfirsthead
\caption{continued.}\\
\hline
ID   &BGPS clumps     &l    &b     &\multicolumn{2}{c}{distance}  &$T_{dust}$  & \multicolumn{2}{c}{R}      &$\rm S_{40}$   &$\rm S_{int}$ &$\rm N_{H_2}$ &$\rm n_{H_2}$  &$\rm M_{iso}$       &$\rm M_{vir}$   &classification          \\
     &                &deg  & deg &kpc &ref                      &K           & arcsec  &pc                & Jy            & Jy           & $\rm\times 10^{22}\,cm^{-2}$      &$\rm\times10^3\,cm^{-3}$       & $\rm M_{\odot}$      &$\rm M_{\odot}$   &                                 \\

(1)  & (2)                  &(3) &(4)                      &(5)         & (6)     &(7)            &(8)           &(9)           &(10)               &(11)           &(12)     &(13)     &(14) &(15) &(16)\\
\hline
\endhead
\hline
\endfoot
896  & G003.254+00.410  & 3.254        & 0.41      &10.49 &3   &13.5    &66.2	  &3.4  &0.34(0.04)    &1.65(0.15)    &1.7(0.2)  & 0.38(0.03)   &4207.2(382.5)      &5633.2(1691.7)   &Quiescent          \\
920  & G003.310-00.402  & 3.310        & -0.402    &2.93  &3   &15.5    &71.84	  &1.0  &0.85(0.06)    &4.22(0.3)     &3.5(0.2)  & 2.21(0.16)   &680.8(48.4)        &1863.3(210.5)    &Protostellar       \\
1129 & G005.833-00.511  & 5.833        & -0.511    &3.43  &1   &19.8    &71.71	  &1.2  &0.77(0.07)    &3.95(0.31)    &2.2(0.2)  & 1.26(0.10)   &617.2(48.4)        &773.4(122.3)     &H II region        \\
1250 & G006.919-00.225	& 6.919        & -0.225    &2.97  &3   &13.1    &67.93	  &1.0  &0.83(0.07)    &4.25(0.32)    &4.4(0.4)  & 3.36(0.25)   &910.6(68.6)        &1516.5(170.8)    &H II region        \\
1290 & G007.337-00.015	& 7.337        & -0.015    &13.55 &3   &15.2    &65.67	  &4.3  &0.51(0.05)    &2.47(0.21)    &2.2(0.2)  & 0.38(0.03)   &8772.9(745.9)      &3373.3(260.2)    &Protostellar       \\
1337 & G008.141+00.224	& 8.142        & 0.224     &2.99  &3   &26.2    &80.81	  &1.2  &2.75(0.19)    &14.24(0.95)   &5.5(0.4)  & 2.52(0.17)   &1170.3(78.1)       &1358.2(134.9)    &H II region        \\
1354 & G008.352-00.318	& 8.352        & -0.318    &4.57  &1   &27.4    &81.4    &1.8  &0.54(0.07)    &2.68(0.29)    &1.0(0.1)   & 0.29(0.03)   &486.2(52.6)        &1082.2(173.1)    &H II region        \\
1412 & G009.212-00.202	& 9.212        & -0.202    &4.56  &1   &16.1    &64.29	  &1.4  &1.02(0.09)    &4.6(0.36)     &4.0(0.4)  & 2.05(0.16)   &1700.1(133.0)      &1803.5(231.4)    &H II region        \\
1416 & G009.282-00.148	& 9.282        & -0.148    &4.51  &1   &11.0    &72.32	  &1.6  &0.32(0.06)    &1.95(0.25)    &2.3(0.4)  & 1.12(0.14)   &1282.8(164.5)      &1162.7(218.7)    &Protostellar       \\
1455 & G010.152-00.344	& 10.152       & -0.344    &1.26  &3   &43.0    &85.98	  &0.5  &1.92(0.14)    &10.33(0.71)   &2.1(0.2)  & 1.98(0.14)   &82.8(5.7)          &1192.0(152.6)    &H II region        \\
1466 & $\ast$G010.214-00.324	& 10.214       & -0.324    &3.12  &3   &24.3    &84.6    &1.3  &1.56(0.13)    &8.3(0.63)     &3.5(0.3)   & 1.35(0.10)   &817.8(62.1)        &5010.0(631.5)    &Protostellar       \\
1467 & G010.226-00.208	& 10.226       & -0.208    &3.13  &3   &18.2    &98.76	  &1.5  &1.00(0.08)    &7.9(0.57)     &3.3(0.3)  & 1.19(0.09)   &1154.7(83.3)       &2033.4(120.3)    &Protostellar       \\
1474 & G010.286-00.120  & 10.286       & -0.12     &3.55  &1   &24.0    &68.12	  &1.2  &1.71(0.14)    &9.41(0.68)    &3.9(0.3)  & 2.62(0.19)   &1219.8(88.1)       &2251.3(314.8)    &H II region        \\
1479 & G010.320-00.258  & 10.32        & -0.258    &3.05  &3   &21.4    &62.32	  &0.9  &0.86(0.07)    &3.36(0.27)    &2.3(0.2)  & 1.65(0.13)   &373.8(30.0)        &1826.8(167.6)    &H II region        \\
1483 & G010.343-00.144  & 10.344       & -0.144    &3.55  &1   &25.3    &84.87	  &1.5  &1.48(0.13)    &8.58(0.61)    &3.1(0.3)  & 1.15(0.08)   &1039.2(73.9)       &1829.3(100.4)    &Protostellar       \\
1507 & G010.625-00.338  & 10.625       & -0.338    &5.2   &1   &22.7    &86.4    &2.2  &1.12(0.10)    &5.45(0.44)    &2.7(0.2)   & 0.55(0.04)   &1630.0(131.6)      &3249.6(244.9)    &H II region        \\
1528 & G010.727-00.332  & 10.727       & -0.332    &4.9   &2   &18.8    &66.59	  &1.6  &0.50(0.05)    &2.28(0.2)     &1.6(0.2)  & 0.68(0.06)   &780.7(68.5)        &1036.4(269.8)    &Protostellar       \\
1569 & G010.999-00.370  & 10.999       & -0.370    &16.57 &3   &16.4    &61.06	  &4.9  &0.27(0.04)    &1.05(0.13)    &1.0(0.2)  & 0.15(0.02)   &4988.5(617.6)      &2298.6(497.9)    &uncertain          \\
1580 & G011.063-00.096  & 11.063       & -0.096    &3.38  &2   &12.2    &77.96	  &1.3  &0.34(0.05)    &2.63(0.23)    &2.0(0.3)  & 1.36(0.12)   &817.9(71.5)        &459.7(106.4)     &Quiescent          \\
1590 & G011.111-00.398  & 11.111       & -0.398    &5.2   &1   &19.9    &81.87	  &2.1  &1.19(0.09)    &7.43(0.52)    &3.4(0.3)  & 1.04(0.07)   &2649.9(185.5)      &2989.1(306.6)    &H II region        \\
1592 $\ast$& G011.121-00.128  & 11.121       & -0.128    &3.03  &3   &11.0    &68.78	  &1.0  &0.44(0.05)    &2.67(0.22)    &3.1(0.4)  & 2.66(0.22)   &792.8(65.3)        &870.1(143.0)     &Protostellar       \\
1684 $\ast$& G012.215-00.118(a)  & 12.208    & -0.104    &14    &1   &26.85   &43.1     &2.9  &4.21(0.91)    &4.21(0.91)    &8.2(1.8)  & 1.02(0.22)   &7353.6(1589.5)     &13748.3(3030.1)  &H II region        \\
1684 & G012.215-00.118(b)  & 12.204    & -0.126    &14    &1   &25.1    &15.9     &1.1  &1.88(0.12)    &1.88(0.12)    &4.0(0.3)  & 9.84(0.63)   &3577.6(228.4)      &3155.5(338.1)    &Protostellar        \\
1742 & G012.627-00.016  & 12.627       & -0.016    &13.45 &3   &18.0    &78.87	  &5.1  &1.15(0.09)    &5.71(0.41)    &3.8(0.3)  & 0.40(0.03)   &15652.1(1123.9)    &10082.5(1104.6)  &H II region        \\
1747 & G012.681-00.182  & 12.681       & -0.182    &10.73 &3   &20.6    &86.15	  &4.5  &2.09(0.14)    &11.82(0.79)   &5.8(0.4)  & 0.66(0.04)   &17127.9(1144.8)    &6089.7(839.4)    &Protostellar       \\
1758 & G012.721-00.216  & 12.721       & -0.216    &3.46  &1   &25.3    &65.87	  &1.1  &0.96(0.09)    &3.81(0.34)    &2.0(0.2)  & 1.12(0.10)   &438.4(39.1)        &1876.2(340.2)    &H II region        \\
1771 & G012.773+00.334  & 12.773       & 0.334     &2.29  &1   &23.1    &66.76	  &0.7  &1.04(0.09)    &5.43(0.43)    &2.5(0.2)  & 2.61(0.21)   &307.8(24.4)        &216.0(51.4)      &H II region        \\
1778 & G012.805-00.318  & 12.805       & 0.318     &1.83  &3   &18.4    &72.81	  &0.6  &0.66(0.07)    &3.17(0.3)     &2.1(0.2)  & 2.00(0.19)   &156.0(14.8)        &421.3(52.9)      &H II region        \\
1809 & G012.905-00.030  & 12.905       & -0.030    &4.6   &1   &19.4    &65.04	  &1.5  &1.03(0.08)    &3.56(0.27)    &3.1(0.2)  & 1.17(0.09)   &1028.7(78.0)       &1471.6(338.1)    &H II region        \\
1810 & G012.909-00.260  & 12.909       & -0.260    &3.06  &3   &22.4    &79.15	  &1.2  &3.17(0.21)    &15.85(1.04)   &7.8(0.5)  & 3.57(0.23)   &1670.6(109.6)      &2258.7(164.2)    &H II region        \\
1871 & G013.217+00.036	& 13.217       & 0.036     &4.36  &1   &20.9    &74.1	  &1.6  &0.96(0.09)    &6.87(0.51)    &2.6(0.2)  & 1.45(0.11)   &1612.0(119.7)      &1892.0(158.9)    &PDR                \\
1876 & G013.245-00.084	& 13.245       & -0.084    &3.59  &1   &12.3    &62.1	  &1.1  &0.94(0.08)    &3.54(0.29)    &5.6(0.5)  & 3.36(0.27)   &1225.6(100.4)      &1453.9(185.0)    &Protostellar       \\
1883 & G013.275-00.336	& 13.275       & -0.336    &3.83  &1   &12.6    &83.71	  &1.6  &0.49(0.05)    &4.04(0.31)    &2.8(0.3)  & 1.41(0.11)   &1531.4(117.5)      &943.0(278.9)     &Protostellar       \\
2011 & G014.194-00.193	& 14.194       & -0.193    &3.63  &1   &15.45   &89.07	  &1.6  &1.40(0.11)    &8.32(0.6)     &5.8(0.5)  & 1.86(0.13)   &2070.1(149.3)      &2575.3(109.4)    &Protostellar       \\
7528 & G349.798+00.108	&349.798       & 0.108     &10.72 &3   &20.9    &73.21	  &3.8  &0.50(0.06)    &2.87(0.26)    &1.4(0.2)  & 0.26(0.02)   &4071.1(368.8)      &5078.3(972.2)    &Protostellar       \\
7549 & G350.177+00.014	& 350.178      & 0.014     &10.38 &3   &18.0    &81.73	  &4.1  &0.96(0.08)    &5.72(0.43)    &3.2(0.3)  & 0.46(0.03)   &9338.6(702.0)      &10890.6(1136.0)  &H II region        \\
7571 & G350.521-00.350	& 350.522      & -0.35     &3.56  &1   &19.2    &61.21	  &1.1  &0.62(0.07)    &2.65(0.26)    &1.9(0.2)  & 1.36(0.13)   &465.2(45.6)        &1095.2(108.3)    &Protostellar       \\
7628 & G351.555+00.206	& 351.556      & 0.206     &11.80 &3   &29.1    &73.84	  &4.2  &2.49(0.18)    &11.05(0.8)    &4.4(0.3)  & 0.57(0.04)   &12398.9(897.7)     &25448.7(7049.6)  &H II region        \\
7714 & G352.584-00.184	& 352.584      & -0.184    &5.1   &2   &23.2    &71.9     &1.8  &0.85(0.10)    &4.94(0.44)    &2.0(0.2)  & 0.85(0.08)   &1381.2(123.0)      &2080.7(161.4)    &PDR                \\
7731 & G352.970+00.358	& 352.97       & 0.358     &1.09  &2   &18.35   &66.07	  &0.3  &0.41(0.09)    &2.72(0.33)    &1.3(0.3)  & 3.87(0.47)   &47.7(5.8)          &1273.2(437.9)    &Quiescent          \\
7748 & G353.091+00.446	& 353.092      & 0.446     &0.73  &2   &19.7    &60	  &0.2  &0.79(0.13)    &4.35(0.44)    &2.3(0.4)  &11.19(1.13)   &31.0(3.1)          &136.1(16.1)      &H II region        \\
7749 & G353.117+00.366	& 353.118      & 0.366     &0.74  &1   &22.8    &83.84	  &0.3  &0.86(0.11)    &5.82(0.54)    &2.1(0.3)  & 4.45(0.41)   &35.0(3.3)          &128.4(24.0)      &Protostellar       \\
7758 & G353.316-00.256	& 353.316      & -0.256    &3.11  &1   &14.2    &85.96	  &1.3  &0.66(0.11)    &4.63(0.49)    &3.1(0.5)  & 1.52(0.16)   &959.9(101.6)       &1545.9(619.9)    &Protostellar       \\
7771 & G353.412-00.360	& 353.412      & -0.36     &3.8   &1   &22.5    &70.35	  &1.3  &11.16(0.7)    &47.62(3.01)   &27.4(1.7) &12.22(0.77)   &7694.8(486.4)      &3212.2(157.4)    &H II region        \\
7794 & G353.978+00.260	& 353.978      & 0.26      &19.76 &3   &12.0    &97.81	  &9.4  &0.52(0.08)    &5.03(0.48)    &3.2(0.5)  & 0.23(0.02)   &54925.5(5241.4)    &6917.9(1475.5)   &Protostellar       \\
7806 & G354.208-00.036	& 354.208      & -0.036    &11.69 &3   &26.3    &62.71	  &3.6  &0.73(0.10)    &2.75(0.32)    &1.5(0.2)  & 0.26(0.03)   &3437.9(400.0)      &6514.8(814.3)    &H II region        \\
7832 & G354.600+00.474(a)& 354.6       & 0.470     &4.2   &1   &24.8    &38.26	  &0.8  &3.58(0.54)    &3.58(0.54)    &7.7(1.2)  & 4.55(0.69)   &622.7(93.9)        &1091.8(92.4)     &H II region        \\
7832 & G354.600+00.474(b)& 354.617     & 0.470     &4.2   &1   &24.1    &25.06	  &0.5  &3.89(0.59)    &3.89(0.59)    &8.7(1.3)  & 18.27(2.77)  &702.0(106.5)       &792.4(115.4)     &H II region        \\
7836 & G354.662+00.484	& 354.662      & 0.484     &12.59 &1   &24.2    &63.53	  &3.9  &1.09(0.16)    &5.26(0.58)    &2.4(0.4)  & 0.50(0.06)   &8484.3(935.5)      &2678.9(263.6)    &H II region        \\
7874 & G355.268-00.270	& 355.268      & -0.27     &1.29  &1   &14.7    &61.07	  &0.4  &1.06(0.10)    &4.19(0.39)    &4.7(0.4)  & 8.80(0.82)   &141.8(13.2)        &849.0(67.5)      &Protostellar       \\

      \hline
\end{longtable}
 $^{\rm a}$ Sources are named by galactic coordinates of BGPS sources. An $\ast$ indicates infall candidates.
References-Distance: (1) \citet{He15}, (2)\citet{He16}, (3)this paper.\\
NOTE: The column are as follows: (1) clump ID; (2) clump names; (3)-(4) the coordinations of the peaks; (5) distance; (6) references for the distances; (7) dust temperature; (8)-(9) the effective physical radius in units of arcsecond and pc, respectively; (10) $40''$ aperture flux density ; (11) integrated flux density; (12) column density; (13) volume density; (14) clump mass derived from the integrated 1.1 mm emission; (15) viral mass; and (16) Spitzer classification.
}
\end{landscape}

\begin{table}
\begin{center}
   \tabcolsep 4.5mm\caption{The whole detection rates of the detected species along with their corresponding detection rates in each evolutionary stage.}
   \label{tab2}
   \small
\def\temptablewidth{10\textwidth}%
\begin{tabular}{lllllllllr}
     \hline\noalign{\smallskip}
\multicolumn{1}{c}{Detected species} & \multicolumn{1}{c}{total detection rate}   & \multicolumn{3}{c}{detection rates in each evolutionary stage} \\
                                     &                                            &  pre-steller   & proto-stellar    & HII/PDR   \\ \hline

N$_2$H$^+$(1-0)               & 100$\%$(50)   &  100$\%$(3)   &  100$\%$(19)   &  100$\%$(27)          \\
HNC(1-0)                      & 100$\%$(50)   &  100$\%$(3)   &  100$\%$(19)   &  100$\%$(27)          \\
HCO$^+$ (1-0)                 & 98.0$\%$(49)  &  100$\%$(3)   &  100$\%$(19)   &  96.3$\%$(26)         \\
HCN (1-0)                     & 86.0$\%$(43)  &  100$\%$(3)   &  94.7$\%$(18)  &  85.2$\%$(23)         \\
HN$^{13}$C (1-0)              & 30.0$\%$(15)  &  0(0)         &  36.8$\%$(7)   &  29.7$\%$(8)          \\
H$^{13}$CO$^+$ (1-0)          & 52.0$\%$(26)  &  33.3$\%$(1)  &  47.4$\%$(9)   &  59.3$\%$(16)         \\
C$_2$H (1-0)                  & 66.0$\%$(33)  &  0(0)         &  63.2$\%$(12)  &  77.8$\%$(21)          \\
HC$_3$N (10-9)                & 46.0$\%$(23)  &  33.3$\%$(1)  &  42.1$\%$(8)   &  51.9$\%$(14)          \\
SiO (2-1)                     & 10.0$\%$(5)   &  33.3$\%$(1)  &  0.05$\%$(1)   &  11.1$\%$(3)           \\
HNCO ($4_{4,0}-3_{0,3}$)      & 0.02$\%$(1)   &  33.3$\%$(1)  &  0(0)          &  0(0)                  \\
$^{13}$CS (2-1)               & 0.02$\%$(1)   &  0(0)         &  0(0)          &  0.04$\%$(1)           \\

      \noalign{\smallskip}\hline
      \end{tabular}
      \\NOTE: Quantities in parentheses represent the detected numbers of the percentage.
      \end{center}
      \end{table}

\clearpage
\setlength{\LTleft}{0pt} \setlength{\LTright}{0pt}
{\scriptsize
 \renewcommand{\footnoterule}{}
  \begin{longtable}{lllllllll}
    \caption{\label{tab3}The derived line parameters of the observed sources. Quantities in parentheses give the uncertainties. }\\
      \hline
BGPS clumps      &range                &lines        &$\rm T_{mb}$   &$\rm\int{T_{mb}dV}$     &  $\rm V_{LSR}$         &$\Delta V$    \\
                 &$\rm(km\, s^{-1})$   &             & (K)           & $\rm(K \,km \,s^{-1})$ & $\rm(km \,s^{-1})$     &$\rm(km\, s^{-1})$ \\

   \hline
   \endfirsthead
\caption{continued.}\\
   \hline
BGPS clumps      &range$\rm(km\, s^{-1})$   &lines        &$\rm T_{mb}$   &$\rm\int{T_{mb}dV}$     &  $\rm V_{LSR}$         &$\Delta V$    \\
                 &$\rm(km\, s^{-1})$        &             & (K)           & $\rm(K \,km \,s^{-1})$ & $\rm(km \,s^{-1})$     &$\rm(km\, s^{-1})$ \\
   \hline
   \endhead
   \hline
   \endfoot
G003.254+00.410           & (48.06-79.00)
& $\rm N_2H^+ (1-0)$    & 1.61(0.28)   & 35.29(1.02)   & 63.0(0.37)   & 11.1(1.67)   \\
&& $\rm HNC (1-0)$       & 1.85(0.27)   & 34.62(0.99)   & 62.69(0.30)  & 18.69(0.73)   \\
&& $\rm HCO^+ (1-0)$     & 0.91(0.28)   & 20.85(1.02)   & 50.68(1.08)  & 53.90(2.57)   \\
&& $\rm HCN (1-0)$       & 1.97(0.29)   & 56.01(1.06)   & 55.6(0.47)   & 34.2(1.80)   \\
&& $\rm SiO (2-1)$       & 0.96(0.28)   & 23.37(1.02)   & 62.52(0.80)  & 28.30(2.01)   \\
&& $\rm HC_3N (10-9)$    & 1.06(0.27)   & 14.14(0.99)   & 62.58(0.41)  & 12.25(1.13)   \\
&& $\rm HNCO (4_{4,0}-3_{0,3})$     & 0.71(0.22)    & 9.36(0.80)   & 62.16(0.73)  & 13.67(2.07) \\

G003.310-00.402           & (2.45-9.28)
& $\rm N_2H^+ (1-0)$    & 1.91(0.43)   & 8.74(0.72)     & 5.59(0.12)   & 3.46(0.20)    \\
\hline

\end{longtable}
 $^a$ There are two velocity ranges associated with the clump. The latter is for HCN(1-0) and the former is for the other detected spectra. $\ast$ indicates an infall candidate.\\ NOTE: The columns are as follows: (1) clump name; (2) the integrated velocity interval; (3) the detected lines in each clump; (4) the main bright temperature; (5) the integrated intensity; (6) the centroid velocity; (7) line width.
}

\clearpage
\setlength{\LTleft}{0pt} \setlength{\LTright}{0pt}
{\scriptsize
 \renewcommand{\footnoterule}{}
  \begin{longtable}{llllllllllll}
    \caption{\label{tab4}The optical depths of N$_2$H$^+$, HN$^{13}$C, H$^{13}$CO$^+$ and C$_2$H, as well as the column densities of N$_2$H$^+$, HNC, HCO$^+$, C$_2$H and HC$_3$N}\\
      \hline
BGPS clumps         &\multicolumn{4}{c}{opacity}        & &     &\multicolumn{5}{c}{column density} \\
                    &N$_2$H$^+$& HN$^{13}$C  & H$^{13}$CO$^+$ &C$_2$H & &  N$_2$H$^+$ & HNC             & HCO$^+$         &C$_2$H            &HC$_3$N \\
                    &          &      &         &       & & $\times10^{13}$ cm$^{-2}$ & $\times10^{14}$ cm$^{-2}$             & $\times10^{14}$ cm$^{-2}$         &$\times10^{14}$ cm$^{-2}$           &$\times10^{13}$ cm$^{-2}$ \\
   \hline
   \endfirsthead
\caption{continued.}\\
   \hline
BGPS clumps         &\multicolumn{4}{c}{opacity}        & &     &\multicolumn{5}{c}{column density} \\
                    &N$_2$H$^+$& HNC  & HCO$^+$ &C$_2$H & &  N$_2$H$^+$ & HNC             & HCO$^+$         &C$_2$H            &HC$_3$N \\
                    &          &      &         &       & & $\times10^{13}$ cm$^{-2}$ & $\times10^{14}$ cm$^{-2}$             & $\times10^{14}$ cm$^{-2}$         &$\times10^{14}$ cm$^{-2}$           &$\times10^{13}$ cm$^{-2}$ \\
   \hline
   \endhead
   \hline
   \endfoot
      &\multicolumn{9}{c}{HII/PDR}    \\ \hline
G005.833-00.511     &0.96     &0.48   & 0.71    &0.52   & &4.25 (0.27)  &  1.75(0.24)   & 3.75(0.33)    &   4.03( 0.58)     & ...            \\
G006.919-00.225     &0.04     &0.36   & 0.29    &...    & &3.13 (0.11)  &  0.70(0.12)   & 0.38(0.05)    &   ...             & 1.26(0.13)     \\
G008.141+00.224     &0.02     &0.29   & 0.48    &1.30   & &4.86 (0.22)  &  1.00(0.11)   & 1.03(0.26)    &   9.77 ( 0.96)     & 1.37(0.18)     \\
G008.352-00.318     &0.05     &...    & 0.20    &1.03   & &2.88 (0.18)  &  ...          & 0.33(0.04)    &   6.89 ( 0.77)     & 0.78(0.16)     \\
G009.212-00.202     &0.37     &0.45   & 0.27    &0.13   & &3.91 (0.14)  &  1.91(0.23)   & 0.56(0.07)    &   2.54 ( 0.34)     & 0.70(0.16)     \\
G010.152-00.344     &5.15     &...    & 0.49    &0.55   & &41.15(2.07)  &  ...          & 3.13(0.35)    &   20.90( 1.20)     & 1.43(0.23)     \\
G010.286-00.120     &0.51     &...    & ...     &...    & &11.59(0.47)  &  ...          & ...           &   ...             & ...            \\
\hline

\end{longtable}
}

\clearpage
\setlength{\LTleft}{10pt} \setlength{\LTright}{0pt}
{\small
 \renewcommand{\footnoterule}{}
  \begin{longtable}{lllllll}
    \caption{\label{tab5} The abundances of N$_2$H$^+$, HNC, HCO$^+$, C$_2$H and HC$_3$N.}\\
      \hline
BGPS clumps     &  N$_2$H$^+$       & HNC                     & HCO$^+$                   &C$_2$H              &HC$_3$N      \\
                & $10^{-9}$         & $10^{-8}$               & $10^{-8}$                 & $10^{-8}$          & $10^{-10}$     \\

   \hline
   \endfirsthead
\caption{continued.}\\
   \hline
BGPS clumps     &  N$_2$H$^+$       & HNC                     & HCO$^+$                   &C$_2$H              &HC$_3$N      \\
                & $10^{-9}$         & $10^{-8}$               & $10^{-8}$                 & $10^{-8}$          & $10^{-10}$     \\
   \hline
   \endhead
   \hline
   \endfoot
    &\multicolumn{4}{c}{HII/PDR}         \\ \hline
G005.833-00.511   & 1.93 (0.22)       &    0.80(  0.13)         &   1.70( 0.21)             &   1.85( 0.31)      &  ...             \\
G006.919-00.225   & 0.72 (0.07)       &    0.16(  0.03)         &   0.09( 0.01)             &   ...              &   2.87( 0.40)    \\
G008.141+00.224   & 0.88 (0.07)       &    0.18(  0.02)         &   0.19( 0.05)             &   1.78( 0.22)      &   2.49( 0.37)    \\
G008.352-00.318   & 2.88 (0.34)       &   ...                   &   0.33( 0.05)             &   6.89( 1.03)      &   7.76( 1.74)    \\
G009.212-00.202   & 0.97 (0.11)       &    0.48(  0.07)         &   0.14( 0.02)             &   0.62( 0.10)      &   1.76( 0.43)    \\
G010.152-00.344   & 19.60(2.11)       &   ...                   &   1.49( 0.22)             &   9.96( 1.10)      &   6.79( 1.26)    \\
G010.286-00.120   & 2.97 (0.25)       &   ...                   &  ...                      &   ...              &  ...             \\
\hline

\end{longtable}
}

\clearpage
\begin{table}
\begin{center}
   \tabcolsep 4.5mm \caption{The profiles of the possible infall candidates.}
    \label{tab6}
   \small
\def\temptablewidth{12\textwidth}%
    \begin{tabular}{lrrrr}
        \hline\noalign{\smallskip}
        \multicolumn{1}{c}{Clump$^{\rm{a}}$} & \multicolumn{1}{c}{$\rm \delta V$\,}             & \multicolumn{1}{c}{$\rm \delta V$\,}         & \multicolumn{1}{c}{Profile} \\
        \multicolumn{1}{c}{name}             & \multicolumn{1}{c}{$\rm HCO^{+}(1-0)$\,}         & \multicolumn{1}{c}{$\rm HNC(1-0)$\,)}        &         \\
        \multicolumn{1}{c}{(1)}              & \multicolumn{1}{c}{(2)}                          & \multicolumn{1}{c}{(3)}                      & \multicolumn{1}{c}{(4)} \\
        \hline
      $\ast$G010.214-00.324       & -0.59(0.06) & -0.37(0.05) &B,B \\
$\ast$G011.121-00.128       & -1.17(0.18) & -0.43(0.09) &B,B \\
$\ast$G012.215-00.118(a)    & -0.25(0.09) &-0.19(0.08)  &B,N \\

        \noalign{\smallskip}\hline
      \end{tabular}
      \\ $^{\rm{a}}$ Sources are named by galactic coordinates of BGPS sources:  An $\ast$ indicates infall candidates.\\
  NOTE: The HCO$^{+}$(1-0) and HNC(1-0) profiles are evaluated as follows: B denotes a blue profile, R denotes a red profile, and N denotes neither blue nor red.
  Quantities in parentheses give the uncertainties in units of 0.01. The columns are as follows: (1) Clump names; (2) asymmetry of HCO$^{+}$(1-0); (3) asymmetry of HNC(1-0); (4) profile of HCO$^{+}$(1-0) and HNC(1-0)\\
      \end{center}
      \end{table}

\end{document}